\documentstyle[preprint]{aastex}
\begin{document}

\title{The Chemical Composition of Carbon-Rich, Very Metal-Poor Stars: A New Class of Mildly Carbon-Rich Objects Without Excess of Neutron-Capture Elements}

\author{Wako Aoki\altaffilmark{1}, John E. Norris\altaffilmark{2}, Sean G. Ryan\altaffilmark{3}, Timothy C. Beers\altaffilmark{4}, Hiroyasu Ando\altaffilmark{1}}
\altaffiltext{1}{National Astronomical Observatory, Mitaka, Tokyo, 181-8588 Japan; email: aoki.wako@nao.ac.jp, ando@optik.mtk.nao.ac.jp}
\altaffiltext{2}{Research School of Astronomy \& Astrophysics, The Australian National University, Private Bag, Weston Creek Post Office, Canberra, ACT 2611, Australia; email: jen@mso.anu.edu.au}
\altaffiltext{3}{Department of Physics \& Astronomy, The Open University, Walton Hall, Milton Keynes, MK7 6AA, UK; email: s.g.ryan@open.ac.uk}
\altaffiltext{4}{Department of Physics \& Astronomy, Michigan State University, East Lansing, MI 48824-1116; email: beers@pa.msu.edu}

\begin{abstract} 

We report on an analysis of the chemical composition of five carbon-rich, very
metal-poor stars based on high-resolution spectra.  One star, CS~22948-027,
exhibits very large overabundances of carbon, nitrogen, and the neutron-capture
elements, as found in the previous study of Hill et al..  This result may be
interpreted as a consequence of mass transfer from a binary companion that
previously evolved through the asymptotic giant branch stage. By way of
contrast, the other four stars we investigate exhibit {\it no} overabundances
of barium ([Ba/Fe]$<0$), while three of them have mildly enhanced carbon and/or
nitrogen ([C+N]$\sim$+1).  We have been unable to determine accurate carbon and
nitrogen abundances for the remaining star (CS~30312-100).  These stars are
rather similar to the carbon-rich, neutron-capture-element-poor
star CS~22957-027 discussed previously by Norris et al., though the carbon
overabundance in this object is significantly larger ([C/Fe]$=+2.2$).  Our
results imply that these carbon-rich objects with ``normal'' neutron-capture
element abundances are not rare among very metal-deficient stars.  One possible
process to explain this phenomenon is as a result of helium shell flashes near
the base of the AGB in very low-metallicity, low-mass ($M\lesssim 1M_{\odot}$)
stars, as recently proposed by Fujimoto et al..

The moderate carbon enhancements reported herein ([C/Fe]$\sim +1$) are similar
to those reported in the famous {\it r}-process-enhanced star CS~22892-052. We
discuss the possibility that the same process might be responsible for this
similarity, as well as the implication that a completely independent phenomenon
was responsible for the large {\it r}-process enhancement in CS~22892-052.

\end{abstract}
\keywords{nuclear reactions, nucleosynthesis -- stars: abundances -- stars: AGB and post-AGB -- stars: carbon -- stars: Population II}

\section{INTRODUCTION}\label{sec:intro}

Many observational and theoretical studies have been undertaken to clarify the
formation and chemical evolution of the Galaxy, as well as the nature of
stellar evolution at the earliest times.  For this purpose, extensive surveys
of metal-deficient stars have been carried out \citep[e.g., the HK survey of
][]{beers92,beers99}, which have identified objects as metal-poor as
[Fe/H]$=-4$ \footnote{[A/B]$\equiv\log(N_{\rm A}/N_{\rm B})$ $-\log(N_{\rm
A}/N_{\rm B})_{\odot}$, and $\log\epsilon_{\rm A}$ $\equiv\log(N_{\rm A}/N_{\rm
H})+12$ for elements A and B}.  One unexpected result of the HK survey is that
many ($\sim$ 10-15\%) of the most metal-poor stars exhibit anomalously strong
CH bands, most easily understood in terms of an excess of carbon in their
atmospheres.  Moreover, the trend of carbon overabundances appears to increase
with declining metallicity \citep{rossi99}, that is, the most metal-deficient
stars exhibit the highest level of carbon overabundances.

Recently, high-resolution abundance studies have been extended to include these
carbon-rich metal-poor stars, but the number of such studies is still quite
limited.  Given that stars with carbon excesses are numerous among the most
metal-deficient stars, they take on considerable importance in the study of the
chemical evolution of the Galaxy. For instance, a recent study of chemical
enrichment in the early Galaxy \citep{abia01} suggests that the initial mass
function at zero metallicity must peak at intermediate stellar mass in order to
account for the carbon and nitrogen enhancement in extremely metal-poor stars.

One explanation for the enhancement of carbon in some of the very metal-poor
stars is that put forward to account for the moderately metal-poor classical CH
stars ([Fe/H]$\sim -$1.5). These are usually explained by a model involving
mass transfer from a carbon-enhanced asymptotic giant branch (AGB) star (one
that has since evolved to the white dwarf stage and cannot now be seen) to its
lower-mass companion, which is now observed as a CH star \citep{mcclure90}. In
this case a carbon-rich metal-deficient star should exhibit the chemical
composition of AGB stars (i.e., evolved intermediate-mass stars) at the
earliest times. One distinct feature of these stars is the enhancement of the
{\it s}-process elements.  The most well-studied example of the phenomenon is
LP~625-44, for which the abundances of 16 heavy elements, including three
elements at the $s$-process peaks (Sr, Ba, and Pb), have been determined
\citep{aoki00}. The variation of radial velocity shown by that work confirms
the binarity of this object, and strongly supports the above mass-transfer
scenario. However, this analysis revealed a quite different abundance trend
from previous theoretical expectations, in particular concerning the
Pb/Ba abundance ratio. In metal-poor AGB stars, the $s$-process was expected to
build up lead, due to the paucity of seed nuclei compared to the (primary)
neutrons, giving a high Pb/Ba ratio \citep[e.g., ] []{gallino98}. However, this
picture is modified on a star-by-star basis by variations in the concentration
of $^{13}$C from which the neutrons originate \citep{busso99}, giving rise to a
range of possible Pb/Ba values \citep{ryan01}.  \citet{aoki01} showed that the
abundance pattern of LP~625-44 is well reproduced by a parametric model of the
{\it s}-process with neutron exposure $\tau\sim 0.7$. Modeling of AGB
nucleosynthesis and evolution where such physical conditions are produced will
promote our understanding of the {\it s}-process and AGB evolution at low
metallicity.

On the other hand, one carbon-rich, very metal-poor star, CS~22892-052,
exhibits an extreme excess of $r$-process, instead of $s$-process, elements
\citep{sneden96,sneden00}. The relationship, if any, between carbon production
and $r$-process enhancement is unclear, but the mechanism of carbon enrichment
may well be quite different from the scenario of mass transfer from an {\it
s}-process-rich AGB star.  The only other presently recognized extremely {\it
r}-process-enhanced metal-poor star, CS~31082-001 (Cayrel et al. 2001),
exhibits no excess of carbon, and hence demonstrates that the two phenomena are
not required to be linked to one another. Furthermore, \citet{norris97b}
\citep[see also ][]{boni98} showed that the carbon-enhanced, very metal-poor
([Fe/H]=$-3.4$) star CS~22957-027 exhibits {\it no} enhancement of
neutron-capture elements. These facts indicate that the canonical mass-transfer
scenario cannot generally explain all carbon-rich objects.

Recently, another mechanism for carbon-enrichment was suggested by
\citet{preston01} to account for their observations of a number of
metal-poor subgiants that do not exhibit clear signs of binarity. These authors
reported on a long-term radial velocity program for carbon-enhanced metal-poor
stars, and found that {\it none} of the three carbon-rich, metal-poor subgiants
in their sample exhibited velocity variations exceeding 0.5~ kms$^{\small -1}$
over an 8 year period.  They suggested that these stars might have undergone an
enhanced mixing event at the end of their giant-branch evolution that
``recycled'' them to the base of the subgiant branch, due to increased hydrogen
mixing into their cores, a scenario originally suggested by
\citet{bond74} to explain the subgiant CH stars he discovered. It
should also be noted that the subgiants studied by \citet{preston01} have
enhanced {\it s}-process elements.

Elemental abundance studies of metal-poor carbon-rich stars {\it without}
excesses of $s$-process elements are quite limited, and the
origin of their carbon enhancement is still unclear. One recent hypothesis for
the formation of these stars is that of \citet{fujimoto00}, who suggested that
production of large carbon and nitrogen excesses may follow from extensive
mixing in low-mass ($M \lesssim 1M_{\odot}$), very metal-poor stars at their
helium core flash at the end of the giant branch, or during the helium-shell
flash phase on the lower AGB.

Thus the carbon-rich, very metal-poor stars show a wide variety of elemental
abundance patterns, and several alternative explanations of their
carbon-enrichment must be considered, in particular for objects without
excesses of neutron-capture elements. In the present paper we report an
abundance analysis based on high-resolution spectroscopy for five carbon-rich,
very metal-poor stars. In \S \ref{sec:obs} we summarize the observations,
report measurements of equivalent widths for detectable spectral lines, and
discuss the determination of radial velocities.  Estimation of the atmospheric
parameters for our program stars is described in \S \ref{sec:param}, where
the effect of carbon and nitrogen excesses on the determination of effective
temperatures is also discussed.  The analysis and results are provided in
\S \ref{sec:ana}. In \S \ref{sec:disc} we summarize the
characteristic chemical compositions of our program stars and discuss possible
interpretations of these results.

\section{OBSERVATIONS AND MEASUREMENTS}\label{sec:obs}

We selected our program sample from the lists of metal-poor stars verified by
medium-resolution spectroscopic follow-up of candidates from the HK survey
\citep{beers92, beers99}, taking into account their brightness, metallicity,
and apparent level of carbon enhancement.  We used the $GP$ index, as defined
in Table 2 of \citet{beers92}, as an indicator of carbon enhancement. The $GP$
index is a pseudo equivalent width of the CH molecular band around 4300~{\AA}
(G band) measured in the medium-resolution spectra.  Our program stars were
taken to be those objects with $GP$ indices that were at least twice the
typical values of other metal-poor stars from the HK survey with similar
metallicity and colors.  Our sample is listed in Table \ref{tab:obs}, along
with additional information concerning the observations.  We note that the $GP$
indices of CS~30314-067 and CS~29502-092 were newly measured from more recently
obtained medium-resolution spectra.

\subsection{Observations and Data Reduction}\label{obs:obs}

Observations were made with the University College London coud\'e
\'echelle spectrograph (UCLES) at the Anglo-Australian 3.9m Telescope during
the period 1994-1998. The observing log is given in Table \ref{tab:obs}.  For
most observations, the Tektronix 1024$\times$1024 CCD was used as the detector;
the MIT-LL 2048$\times$4096 CCD was used during the 1998 April run.  The
spectra were taken with a resolving power $R \approx 40,000$.  Column (1) of
Table \ref{tab:obs} lists the star name.  Columns (2) and (3) list the $V$
apparent magnitude and $B-V$ color reported in previous HK survey papers. The
$GP$ index for each star is listed in column (4).

Data reduction was accomplished using standard procedures within the
IRAF\footnote{IRAF is distributed by the National Optical Astronomy
Observatories, which is operated by the Association of Universities for
Research in Astronomy, Inc. under cooperative agreement with the National
Science Foundation.} suite, including bias subtraction, flat-fielding,
background subtraction, and wavelength calibration.  Cosmic-ray removal was
done by comparing two or more frames obtained during different exposures for
each star, using the same spectrograph setup.  The wavelength intervals,
exposure times, and numbers of exposures (in parentheses) for each of our
program stars are listed in columns (5) and (6) of Table \ref{tab:obs}.

Examples of the spectra in the wavelength region around 4550{\AA} are shown in
Figure \ref{fig:sp4554}, along with identifications of the prominent absorption
lines. The absorption lines of the heavy elements, especially the
\ion{Ba}{2} $\lambda$4554 feature, are discussed in the following
sections. For comparison, the spectrum of an {\it s}-process-rich, metal-poor 
star, LP~625-44, on which we reported separately \citep{aoki00,aoki01}, is
also shown in the Figure.

In order to assess the quality of the data, we constructed a summed spectrum
and a difference spectrum from two spectra obtained by different exposures for
each object. The signal-to-noise (S/N) ratio achieved was based on the signal
level of the sumned spectrum for the nearly line-free region near the peak of
\'echelle blaze profile, compared with the standard deviation of the difference
spectrum for the same region. Column (7) of Table \ref{tab:obs} lists the S/N
ratios at 4350{\AA} for the blue setup, and those at 6000{\AA} for the red one.
These values are comparable to the square root of the number of photons
obtained at each of the listed wavelengths.  

\subsection{Measurement of Equivalent Widths}\label{obs:mes}

Equivalent widths of the extracted and wavelength-calibrated spectra were
measured by fitting Gaussian profiles to the absorption lines. The local
continuum levels were estimated interactively during the fitting procedure.  In
order to reject lines that may suffer from contamination by other features, we
checked the atomic line list provided by \citet{kurucz95}, and the line list of
CH and CN molecules used in the present work (see Section \ref{sec:cn}). The
equivalent widths measured for individual atomic lines are listed in Table
\ref{tab:ew}.

Random (internal) errors in the equivalent widths were estimated by comparing
two or more measurements of those lines in the spectra obtained with
individual exposures. The standard errors for lines with equivalent widths
smaller than 60 m{\AA} are provided in column (8) of Table
\ref{tab:obs} ($\sigma_{\rm W}$).  These internal errors will be used
below for estimation of abundance uncertainties, and for estimation of upper
limits on abundances for elements with no detectable spectral lines.

We checked the dependence of $\sigma_{\rm W}$ on line strength for
CS~30314-067 and CS~29502-092, in which a number of strong and weak
lines lines were measured. The discrepancy between the $\sigma_{\rm W}$ derived
from the weak lines ($W<60$m{\AA}) and that derived from stronger lines is
neither significant nor systematic: $\Delta\sigma_{\rm W}$(strong $-$ weak)
=$-$0.9~m{\AA} and +0.3~m{\AA} for CS~30314-067 and CS~29502-092, respectively.
We note that the uncertainty in the derived elemental abundance (at fixed
$\sigma_{\rm W}$) for strong lines is generally larger than that obtained for
weak lines, due to the saturation effect on line absorption.  

The estimated $\sigma_{\rm W}$ is dependent on the S/N ratio of the spectrum at
each given wavelength; the S/N ratio at the edge of the \'{e}chelle blaze
function is lower than at the center. However, the free spectral range is
well-covered for the blue region, in which most of the lines used in the
present analysis are found, and the S/N ratio is not severely dependent on the
line position in the \'{e}chelle order after combining the spectra for
neighboring orders.  On the other hand, the S/N ratio of the spectra in the
bluest regions is considerably lower than at intermediate wavelengths.  We have
calculated $\sigma_{\rm W}$ for lines with $\lambda <$4000{\AA} for
CS~30314-067 and CS~29502-092, in which many iron lines with $\lambda
<4000${\AA} were detected. The $\sigma_{\rm W}$ derived is roughly 2~m{\AA}
higher than that obtained from lines in the longer wavelength regions. This
indicates that we underestimate the uncertainty in the resulted abundances (by
a maximum of 50\%) for some elements (e.g., Mg, Ti and Fe) for which the
analysis includes lines with $\lambda <$4000{\AA} .  However, the error arising
from $\sigma_{\rm W}$ is not the dominant factor in the total uncertainty in
the resulting abundances for these elements (see section \ref{ana:error}),
hence we simply adopted the $\sigma_{\rm W}$ given in Table 1 for the purpose
of error estimation.
 
\subsection{Measurement of Radial Velocities}\label{obs:rv}

Some of the carbon-rich, very metal-poor stars may be accounted for by mass
transfer across a binary system, as mentioned in section \ref{sec:intro}.
Hence, a search for radial velocity variation is quite important for a test of
this scenario.   Since we have observed each object at only one epoch, we
cannot discuss the variation of radial velocities here, but these measurements
provide a baseline against which future studies can be compared.

For the measurement of radial velocity, we selected clean Fe {\small I} lines
between 4800 and 5400~{\AA} for which the data quality is good, and blending
with CH and CN molecular lines is not severe. The line position was measured by
fitting Gaussian profiles for about 20 lines, then deriving the heliocentric
radial velocity.  Results are provided in column (9) of Table \ref{tab:obs},
along with the standard error in each measurement.  For reference, we also
list the Julian Date (JD) of the observations in column (10) of Table
\ref{tab:obs}. 

Radial velocity variation has already been established for CS~22948-027 by
\citet{preston01}, who also provided orbital parameters (e.g., $P=505$
days) for this object. Our result ($V_{\rm rad}=\;-63.3~$kms$^{\small -1}$) is
in good agreement with the value expected on JD=2449520 from their orbital
parameters, and confirms the binarity of this star.

\section{STELLAR PARAMETERS}\label{sec:param}

In order to perform a detailed abundance analysis using a model-atmosphere
approach, we must first determine the appropriate atmospheric parameters, i.e.,
the effective temperature, surface gravity, metallicity, and micro-turbulence,
to be applied.  Strong molecular absorption due to carbon (and nitrogen) 
can affect the determination of effective temperatures obtained from comparison
with broadband photometry. Below we report a method for determination of
effective temperature that takes these effects into consideration in some
detail (Section \ref{sec:temp}).  The determination of other atmospheric
parameters is discussed in \S \ref{sec:params}.

\subsection{Effective Temperatures for Carbon-Enhanced Stars}\label{sec:temp}

Accurate estimation of effective temperature is crucial for accurate
measurement of elemental abundances. Effective temperatures are frequently
determined from the comparison of broadband colors (e.g., $B-V$) with
temperature scales predicted by model atmospheres \citep[e.g., ][]{bessell98}.
However, for carbon-rich objects the temperature scale of normal (oxygen-rich)
stars cannot necessarily be applied, because of the strong absorption bands of
carbon-bearing molecules, such as CH, CN and C$_{2}$, which can severely affect
the emergent flux distribution.  Strong CH bands, for instance, which are the
defining characteristics of carbon-rich, metal-poor stars such as those in our
program, reduce the flux in the $B$ band, hence increase (redden) the $B-V$
color compared with normal metal-poor stars of the same effective temperature.
In addition to carbon, nitrogen is sometimes found to be overabundant in
carbon-rich stars, and can also affect the emergent flux. For these
reasons, it is necessary to consider an effective temperature scale that
includes the effects of carbon and nitrogen excess of varying amounts.

To evaluate the effect of molecular absorption on the spectra of metal-poor
stars, we calculated synthetic spectra that included molecular lines, as well
as strong atomic lines, in the wavelength region 3000-10000{\AA}. About 20,000
strong Mg, Ca, Ti, and Fe lines were selected from the line list of
\citet{kurucz95}.  Balmer lines of hydrogen were also included in the
calculation, as were molecular lines due to MgH and SiH listed in
\citet{kurucz93b}.  The lines arising from the CH $A^{2}\Delta$-$X^{2}\Pi$ and
$B^{2}\Sigma^{-}$-$X^{2}\Pi$ bands, the CN violet and red systems, and the
C$_{2}$ Swan system, explained in Section \ref{sec:cn}, were used in addition.
The total number of molecular lines is about 55,000.

As a check on the suitability of the synthetic spectra obtained by this
approach, we calculated the spectra for several metallicities with solar
abundance ratios (i.e., [C/Fe]=[N/Fe]=0), using the model atmospheres of
\citet{kurucz93a}, and compared them with Kurucz's emergent fluxes, calculated
taking into account huge numbers of atomic and molecular lines. Though the
number of lines in our list is obviously insufficient to reproduce Kurucz's
emergent fluxes for solar metallicity ([M/H]=0), the fluxes for [M/H]$ \leq-2$
are well reproduced by our synthetic spectra. We conclude that our line list is
adequate to calculate the colors of the metal-poor stars discussed in the
present work.

Synthetic spectra were calculated for several levels of carbon and nitrogen
enhancement, using the above line list, for models with metallicity [M/H] $=-2$
and $-3$, and for $T_{\rm eff}$ ranging from 4000K to 6000K.  Calculations were
done for $T_{\rm eff}=$4000K, 4500K, 5000K, and 5500K for $\log g=2.0$, $T_{\rm
eff}=$5000K, 5500K, and 6000K for $\log g=3.0$, and $T_{\rm eff}=$5500 and
6000K for $\log g=4.0$. The model atmospheres by \citet{kurucz93a} were 
used for this purpose.  Then, fluxes for several photometric bands
\citep[Johnson-Cousins system, ][]{bessell90} were obtained from the synthetic
spectra. The calculated $B-V$ and $R-I$ values are shown as a function of model
effective temperature in Figure \ref{fig:teff}. We show the result for $\log
g$=3.0 for $T_{\rm eff}=$6000K, and those for $\log g$=2.0 for $T_{\rm
eff}=$4000, 4500, 5000 and 5500K in this Figure. 

The effect of surface gravity is not large in these parameter ranges.  The
difference in $B-V$ for $\log g=$3.0 and for $\log g=$4.0 is 0.024 magnitudes
for [Fe/H]$=-3.0$, [C/H]=[N/H]$=-1.0$ and $T_{\rm eff}=$6000K.  For other
cases, the impact of changes in surface gravity on $B-V$ is even less; the
effect on $R-I$ is negligible.

As can be seen in Figure \ref{fig:teff}, $B-V$ is larger (redder), for higher
[C/H] and [N/H], at a given effective temperature. This is because the emergent
flux in the $B$ band is severely affected by CH and CN bands, while the C$_{2}$
absorption in the $V$ band is not strong, except for cool stars ($T_{\rm eff}
\lesssim 4500$K) with extremely high [C/Fe]. On the other hand, $R-I$ is
smaller (bluer) for higher [C/H] and [N/H] at a given effective temperature.
The CN red system is important for the red and near-infrared regions in
carbon-rich stars, and the effect should be larger in the $I$ band than in the
$R$ band. As can be seen in this Figure, the effect of carbon and nitrogen
excess is much smaller in $R-I$ for $T_{\rm eff}\leq5000$K; in such cases the
$R-I$ color is preferable for the determination of effective temperature.  In
the present work, however, only the $B-V$ colors were available for most of the
program stars, hence we make use of the effective temperature scale based on
those values.  Future measurements of $R-I$ colors would be most useful.

In the above calculations, model atmospheres produced for solar elemental
abundance ratios ([X/Fe]=0) were used. However, the effect of carbon and
nitrogen excesses on atmospheric structure should also be considered in the
model calculation, because the absorption due to many molecular lines blocks
the flux, and the atmosphere heats up due to the back-warming effect. In this
sense our calculation is incomplete, in particular for cool objects.  We expect
that this effect will have only a small impact, however, for stars with
effective temperatures in the range considered in the present work.

The model atmospheres and effective temperature scale of metal-poor stars with
carbon and nitrogen excess were also studied by
\citet{hill00}. The $B-V$ and $R-I$ colors provided by \citet{hill00},
as a function of effective temperature, are also shown in Figure
\ref{fig:teff} for the same set of chemical compositions assumed,
i.e., [Fe/H] $=-3$ and [C/H] = [N/H] $=-1$. Since the effect of carbon and
nitrogen excess on the atmospheric structures is taken into account in their
calculation, their result may be more realistic than ours. Nevertheless, our
colors, derived for [Fe/H]$=-3$ and $T_{\rm eff} \gtrsim 4500$K, agree well
with those obtained by \citet{hill00} (see Figure \ref{fig:teff}). This
supports the supposition above that the effect of carbon and nitrogen
enhancement on the atmospheric structure is not large for stars with effective
temperature higher than 4500K. For stars with lower temperatures our scale may
be less than optimal.  We note, however, that the above procedure significantly
improves the estimation of the effective temperatures of carbon-rich stars, and
should be taken into account, especially when $B-V$ is the only available
color.

Typical errors in the reported photometry results in an uncertainty $\Delta
(B-V)$ of about 0.02 magnitudes \citep{beers85}. Several of our program stars
may suffer from interstellar reddening of up to 0.05 magnitudes. The $\Delta
T_{\rm eff}$/$\Delta (B-V)$, in the temperature region of interest, is
160$\sim$200K/0.1 mag.  Hence, a typical uncertainty in the $T_{\rm eff}$
estimation is about 100K. The effect of this uncertainty on elemental abundance
determinations is estimated in \S \ref{ana:error} below.

The effect of carbon and nitrogen excess on the determination of effective
temperatures is quite important for our two coolest stars, CS~30314-067 and
CS~22948-027. At $B-V=1.13$, the reddest color for stars in our sample, the
effective temperature derived for the assumption of [Fe/H]$=-3$ and
[C/H]=[N/H]$=-1$ is higher by about 400K than that obtained when no excess of
carbon or nitrogen is present.  We iterated the determination of effective
temperature and the abundance analysis of carbon and nitrogen for each object
until consistency between the resulting abundance and the assumed one in the
analysis was achieved.

CS~22948-027 was also studied by \citet{preston01}, who derived an effective
temperature including the blanketing effects by CH and CN bands.
Their effective temperature is 4640K, which agrees very well with our
estimation (4600K). Like us, they did not include the backwarming effect on the
model atmosphere. The effective temperature about 4600K derived by the present
study and \citet{preston01} is lower than the result of \citet[][$T_{\rm
eff}=4800$K]{hill00}, who included the backwarming effect in the model
calculation.  This discrepancy may be due to the effect of carbon enhancement
on atmospheric structure.

After the present analysis was completed, we became aware of new results for
near-infrared photometry listed in the 2nd Incremental Release Point Source
Catalog (PSC) from the 2MASS survey for three of our stars\footnote{ This
research has made use of the NASA/IPAC Infrared Science Archive, which is
operated by the Jet Propulsion Laboratory, California Institute of Technology,
under contract with the National Aeronautics and Space Administration.  }.
These new data enable us to estimate the effective temperature from other
colors, such as $V-K$, which are less sensitive to line blanketing as
compared with $B-V$. We applied the temperature scale for $V-K$ produced by
\citet{alonso99}, using the 2MASS $K$ photometry ($K=9.03 \pm 0.03$, $10.03 \pm
0.03$, and $9.60 \pm 0.03$ for CS~30314-067, CS~22877-001, and CS~29502-092,
respectively, where the magnitudes and errors are the ``default'' values
provided in the PSC). The derived effective temperatures are 4320K, 4920K and
4800K, respectively.  These temperatures are systematically lower than those
derived from $B-V$ in the present analysis, although the disagreement is not
severe (80K--200K). We do not pursue this issue farther here, but
investigation of this discrepancy, e.g., from studies of blanketing in the
near-infrared region, will be valuable for more reliable estimation of
effective temperatures for carbon-rich objects in the future.

\subsection{Surface Gravity, Metallicity, and Micro-Turbulence}\label{sec:params}

We began our analysis assuming $\log g=2.0$ for all of our stars.
Then the surface gravities were revised to obtain ionization balance
between \ion{Fe}{1} and \ion{Fe}{2} lines. We assumed LTE
in the parameter estimates as well as in the abundance
analysis. The non-LTE effect is important in warm, low gravity stars,
but is expected to be negligible in subgiants and cool giants studied
here \citep[e.g., ][]{gratton99}.  The final derived surface gravities
are listed in Table \ref{tab:param}.  Note that CS~30312-100 has the
gravity of a subgiant, while the other four stars appear to be giants.
Metal abundances were initially assumed to be [M/H] $=-2.5$, then
revised to the [Fe/H] values derived from the abundance analysis for
\ion{Fe}{1} and \ion{Fe}{2} lines.  These parameters were also revised
once the effective temperature was re-determined, taking the excess of
carbon and nitrogen into account, as described in the above
subsection.

The microturbulent velocities ($v_{\rm tur}$) were determined from the
\ion{Fe}{1} lines by demanding no dependence of derived abundance on
equivalent widths. For CS~22948-027, we also measured the microturbulent
velocity using the CN lines with measurable equivalent widths, and confirmed
the agreement with that determined from the \ion{Fe}{1} lines.  The effect of
microturbulent velocity on the atmospheric structure was neglected, and the
ATLAS9 model atmospheres with $v_{\rm tur}=2$~kms$^{\small -1}$ were used for
the abundance analysis, with the microturbulence reset to the
values listed in Table \ref{tab:param}.

\section{ABUNDANCE ANALYSIS AND RESULTS}\label{sec:ana}

The elemental abundance analysis was performed using the model atmospheres in
the ATLAS9 grid of \citet{kurucz93a}. The standard analysis, based on the
equivalent widths measured in \S \ref{obs:mes}, was obtained for most
species, employing the LTE spectral synthesis code used in \citet{aoki97a}.  As
in \citet{ryan96}, Uns\"{o}ld's (1955) treatment of van der Waal's broadening,
enhanced by a factor of 2.2 in $\gamma$, was used. To check our calculations of
elemental abundances, we analyzed some metallic species in the metal-poor
([Fe/H]$\sim -$2.7) giant HD~122563 \citep{norris96} using Kurucz's LTE code
WIDTH6, as well as with our own software. The results agree well; differences in
derived abundances are generally smaller than 0.02~dex for weak lines ($W< 80
$m{\AA}). There is a systematic discrepancy of derived abundances based on
strong lines between the two analyses ($\lesssim 0.10$~dex), which is likely
due to the differences of the treatment of the line broadening \citep[see
][]{ryan98}. Since most elemental abundances in the present work relied
exclusively on weak lines, the influence on the final results is negligible.
Exceptions are the derived abundances of Na, Al, Si and Sr, which relied on
strong lines.

In addition to the above standard analysis, the spectrum synthesis method
was applied to some metallic lines that are affected by blending. The
abundances of carbon and nitrogen were also determined by spectrum synthesis
of the CH, C$_{2}$ and CN bands. Carbon isotope ratios were derived from
analysis of the CH and CN lines.

The analysis was started for \ion{Fe}{1} and \ion{Fe}{2} lines using the
assumed stellar parameters discussed above.  From the analysis of iron lines,
the microturbulence was determined, and surface gravity and metallicity were
revised. The carbon and nitrogen abundances were derived using these (revised)
stellar parameters. After the abundances of C, N and Fe were tentatively
determined by this procedure, the effective temperature was revised making use
of the effective temperature scale described in \S \ref{sec:param}. This
procedure was iterated until consistency between the derived C and N
abundances, and those assumed for the analysis, was achieved.

\subsection{Atomic Lines}\label{ana:metal}

The standard analysis, based on equivalent widths, was made for most atomic
lines without contamination with other lines. The atomic line data compiled by
\citet{norris96} was used for most elements. We supplemented this list with
additional lines, especially for the wavelength region 3700-3760{\AA}, as well
as for the red spectral regions. In the star CS~22948-027, many lines of
neutron-capture elements have been detected. The line list of neutron-capture
elements produced from the same sources referred to by \citet{aoki01} was
applied to the analysis of this star. The references for the {\it gf}-values
are given in Table \ref{tab:ew}.

The effect of hyperfine splitting on the lines of Mn (4030{\AA}, 4033{\AA} and
4034{\AA}) and Co (3845{\AA}, 3873{\AA}, 3894{\AA} and 4121{\AA}) was taken
into consideration using the estimation procedure shown in Figure 1 of
\citet{ryan96}.

The results of our abundance analysis are listed in Table \ref{tab:res}.  One
important conclusion is that the abundances of heavy neutron-capture elements
($Z\geq 38$) are quite low for four of our stars, in contrast to CS~22948-027,
which exhibits a large excess of these elements.  Figure \ref{fig:sp4554} shows
portions of spectra around \ion{Ba}{2} $\lambda 4554$ for two of the  Ba-poor
stars (CS~30314-067 and CS~29502-092) and the {\it s}-process-rich star
LP~625-44 \citep{aoki00}.  Note that the \ion{Ti}{2} and \ion{Fe}{2} lines for
the two (giant) Ba-poor stars are stronger than those in (the subgiant)
LP~625-44, due to their lower temperatures and gravities.  By way of contrast,
the \ion{Ba}{2} $\lambda 4554$ lines in the two giants are much weaker than
that seen in LP~625-44. No lines of other neutron-capture elements are detected
in the two giants, while numerous lines of \ion{La}{2}, \ion{Ce}{2}, \ion{Nd}{2}
and \ion{Sm}{2} appear in LP~625-44.  Figure \ref{fig:sp4554} clearly shows the
low abundances of neutron-capture elements in the stars studied in the present
work.  This result is discussed in detail in section \ref{sec:disc}.

Our spectrum of CS~22948-027 includes \ion{Eu}{2} $\lambda6645$, and a
Eu abundance [Eu/Fe]$=+1.57$ was derived from this line.  However, this
result is rather uncertain due to possible contamination from molecular
absorption lines. No clear Eu line was detected in the spectra of our other
program stars.  Nevertheless, upper limits on Eu abundance are still useful,
because the Ba/Eu ratio may allow us to distinguish the origin of heavy
elements in these stars, i.e., whether they are to be primarily associated with
the {\it r}-process or with the {\it s}-process.  We estimated upper limits
on Eu abundance for CS~30314-067, CS~29502-092 and CS~22877-001 from the
\ion{Eu}{2} $\lambda$4129, taking 3$\sigma_{\small \rm W}$ at this
wavelength as the upper limit of the equivalent width. The results are listed
in Table \ref{tab:res}.  The upper limits on the Eu abundances for CS~29502-092
and CS~22877-001 are uninteresting, because the lower limits on [Ba/Eu] in
these stars are lower than the Solar System {\it r}-process value \citep[e.g.,
{[Ba/Eu]}$=-0.69$, ][]{arlandini99}. The lower limit of [Ba/Eu] in CS~30314-067
is about $-0.1$, higher than the Solar System {\it r}-process value, though
still much lower than the {\it s}-process value
\citep[e.g., {[Ba/Eu]}=+1.15,][]{arlandini99}. This suggests that there is a
possible contribution of the {\it s}-process to the heavy elements in this
star. We note for completeness that the [Ba/Eu] of CS~22948-027 is about
$+0.3$, which also implies a significant contribution of the {\it s}-process to
the heavy elements in this object.

The abundance patterns of the other metals seem rather typical for normal
(i.e., non carbon- and nitrogen-enhanced) metal-poor stars
\citep{mcwilliam95,ryan96}: the $\alpha$-elements (Mg, Si, Ca, and Ti)
are overabundant by about 0.45 dex on average, while Cr and Mn are
underabundant by 0.4 dex, and Co is overabundant by 0.19 dex.

\subsection{Carbon and Nitrogen}\label{sec:cn}

The carbon abundance was determined by obtaining fits of synthetic spectra of
the CH $A^{2}\Delta$-$X^{2}\Pi$ band at 4323{\AA} to the observed ones for
CS~30314-067, CS~29502-092 and CS~22877-001. It is also possible to determine
the carbon abundance from the CH 4315{\AA} band (G-band), but the uncertainty
would be large due to the difficulty in continuum specification. Therefore we
prefer the CH 4323{\AA} band to the G-band for determination of carbon
abundance. The same analysis with C$_{2}$ Swan bands at 5165{\AA} and 5635{\AA}
was applied to CS~22948-027. We note that our spectrum of CS~30312-100 does not
cover the CH band, and the C$_{2}$ bands are not detected in the spectrum.

The nitrogen abundances were derived by a similar analysis for the CN band
(violet system) at 3880{\AA} for CS~ 30314-067, CS~29502-092 and CS~22877-001,
and by the standard analysis of the absorption lines of the CN red system for
CS~22948-027. To evaluate the carbon isotope ratio, the lines of the CH
$B^{2}\Sigma^{-}$-$X^{2}\Pi$ band were used.

The line lists of these molecular bands were produced not only for the
abundance analysis, but also to check on the blending of molecular lines, and
for the calculation of the effective temperature scale. The CH and CN lines
used in \citet{aoki01} were also applied to this analysis. We also used the
lines of the C$_{2}$ Swan bands and the CN red system studied by
\citet{aoki97a} for the analysis of CS~22948-027. A dissociation energy of 7.75
eV \citep{aoki97b} was adopted in the analysis of CN lines. However, there are
some suggestions that lower values of the CN dissociation energy might apply
\citep[see references in][]{aoki97b}, and the uncertainty on nitrogen
abundances derived from CN lines is large: the error in nitrogen abundance
arising from an error of 0.1 eV in the CN dissociation energy is estimated to
be 0.1-0.15 dex in our stars. We note, however, that the uncertainty is
systematic, and has only a small effect on our discussion of the relative
abundances between the stars studied in this work.  Examples of the spectra at
the CN 3880{\AA} band and the CH 4323{\AA} band are shown in Figure
\ref{fig:chcn}, along with the synthetic spectra for three different
assumed nitrogen and carbon abundances, respectively.

The C$_{2}$ Swan bands in CS~22948-027 spectrum are shown in Figure
\ref{fig:c2}, along with the synthetic spectra for three different carbon
abundances. The carbon abundance determined from the C$_{2}$ 5635{\AA} band
was adopted for CS~22948-027 because of the large uncertainty in
continuum determination for the C$_{2}$ 5165{\AA} band.

The carbon isotope ratios ($^{12}$C/$^{13}$C) for CS~30314-067 and CS~29502-092
were determined from the nearly clean CH lines around 4200{\AA}, adopting the
line data used in \citet{aoki01}.  Examples are shown in Figure
\ref{fig:c1213}. In the upper three panels, we show the observed spectrum of
CS~30314-067 and three synthetic ones calculated for $^{12}$C/$^{13}$C=3, 5,
and 10, respectively, for the three CH lines. We find that a $^{12}$C/$^{13}$C
ratio of about 5 is well determined by this analysis. The lower panels show the
spectrum of CS~29502-092, along with the synthetic ones for
$^{12}$C/$^{13}$C=10, 20, and 40, respectively. In this case, the estimation of
the isotope ratio is difficult, because the $^{13}$CH lines are much weaker
than those in CS~30314-067, hence a higher S/N spectrum is desirable for
accurate determination of the ratio in this star. However, note that the
$^{12}$C/$^{13}$C ratio of CS~29502-092 ($\sim 20$) is obviously higher than
that of CS~30314-067. Only a lower limit ($^{12}$C/$^{13}$C$>10$) is estimated
for CS~22877-001, because of the low quality of the spectrum.  The carbon
isotope ratio of CS~22948-027 ($^{12}$C/$^{13}$C$\sim 10$) was determined using
CN lines around 8000{\AA}, following \citet{aoki97a}.

The $^{12}$C/$^{13}$C ratios for CS~29502-092, CS~22948-027, and possibly for
CS~22877-001, are fall in the range 10-20, which are typical values for
classical CH stars \citep{vanture92, aoki97a}. These values may reflect not
only carbon production by the triple-$\alpha$ reaction, but also from
subsequent CN cycles during the hydrogen burning phase(s) \citep{aoki97a}. The
$^{12}$C/$^{13}$C of CS~30314-067 is substantially lower, near the equilibrium
value for the CN cycle ($^{12}$C/$^{13}$C$\sim$3--4). This result is consistent
with the high nitrogen abundance of this object ([N/Fe]=+1.2) which is also
expected from operation of the CN-cycle.  This probably indicates that the
material in the envelope of this star was strongly affected by hydrogen burning
during its evolution on the red-giant branch. The low derived gravity of this
star ($\log g=0.7$) implies that this object has considerably ascended the
red-giant branch, and supports the above interpretation.

\subsection{Upper Limits on Lithium Abundance}\label{ana:oli}

A wide spread in Li abundances is known to exist among carbon-rich, metal-poor
stars \citep[e.g., ][]{norris97a}, and Li may be a key element to understand
the formation of these objects. Though most of our stars are giants, and Li is
not expected to presently exist on their stellar surfaces due to destruction
from internal mixing processes, it is still worthwhile to search for its
presence.  We inspected our spectra around the Li $\lambda$6707 lines, but did
not detect any features at the wavelength of Li. We estimated the upper limit
on the Li abundance for these objects assuming the 3$\sigma_{\small \rm W}$
(Table \ref{tab:obs}) as the upper limit of the equivalent width.  Note,
however, that \citet{hill00} have derived a Li abundance for CS~22948-027,
log~$\epsilon_{\rm Li}$ = +0.3, at a much lower level than our upper
limit. It should be pointed out that the derived upper limits for our
objects are lower than the primordial Li abundance, which is reasonable for
stars in which internal mixing has affected the surface abundances.

\subsection{Error Estimates}\label{ana:error}

The errors of the abundance analysis for our stars were estimated
following the procedures of \citet{ryan96}. The errors arising from
uncertainties in the atmospheric parameters were evaluated by adding
in quadrature the individual errors corresponding to $\Delta T_{\rm
eff}=100$K, $\Delta \log g=0.3$, and $\Delta v_{\rm
tur}=0.5$~kms$^{\small -1}$ for CS~30314-067 (a giant). The errors for
$\Delta T_{\rm eff}$ and $\Delta \log g$ derived for CS~30314-067 were
then applied to the other three giants, while the errors derived for
LP~625-44 (a subgiant) in \citet{aoki00} were applied to the subgiant
CS~30312-100. The uncertainty $\Delta v_{\rm tur}$ was calculated for
individual objects. Internal errors were estimated by assuming the
errors in the measurement of equivalent widths assessed in \S
\ref{sec:obs}, adopting a random error on the $gf$ values of 0.1
dex. The results for the giant CS~30314-067 and the subgiant
CS~30312-100 are given in Table \ref{tab:error}. We note that the
errors are provided for the abundance ratio ([X/Fe]), instead of
[X/H], except for iron.  The effect of the uncertainty in atmospheric
parameters on an individual elemental abundance ($\Delta$[X/H])
sometimes almost cancels out that obtained for the iron abundance
($\Delta$[Fe/H]), resulting in a quite small uncertainty in
[X/Fe]. Final estimated abundance uncertainties are included in Table
\ref{tab:res}.

For carbon and nitrogen abundances determined by spectrum synthesis, we assumed
a 0.1 dex fitting error, and a 0.1 dex uncertainty in the $gf$ values. The
effect of atmospheric parameters are estimated from the analyses assuming the
above uncertainties. We estimated the uncertainty of the nitrogen abundance
ratio from the CN band taking into account the uncertainty of carbon abundance
derived from the CH band.

\subsection{Comparison with Previous Work}\label{sec:comp}

One of our objects, CS~22948-027, was studied in detail by \citet{hill00}.  Our
results for iron, carbon, and nitrogen abundances agree with their values
within the derived uncertainties. The $^{12}$C/$^{13}$C ratio also agrees
rather well with their result, and the large overabundances of Na and Mg
shown by their work are also confirmed. The large overabundances of
neutron-capture elements for this star are also clearly supported by the
present work.  However, note that there is a systematic difference in the
abundances of heavy elements ($Z\geq39$); our analysis obtains 0.2-0.5 dex
smaller abundances. One reason for this discrepancy is the difference in
atmospheric parameters adopted in the two analyses. Our effective temperature
and gravity ($\log g$) are lower by 200~K and 0.8~dex, respectively, than those
of \citet{hill00}. The abundance ratios of these heavy elements to iron
are not so sensitive to the effective temperature assumed, but they {\it are}
rather sensitive to the adopted gravity. The abundance of Ba increases with
increasing $\log g$, while the Fe abundance derived from \ion{Fe}{1} lines is
insensitive to $\log g$. Our lower $\log g$ results in smaller abundance ratios
of Ba and the other neutron-capture elements compared with those in
\citet{hill00}.  Our larger microturbulent velocity also results in lower
abundances when strong lines (e.g., \ion{Ba}{2} lines in CS~22948-027) are used
in the analysis.

The elemental abundances of CS~22948-027 derived by our analysis agree rather
well with the results of \citet{preston01}, who adopted similar values of
atmospheric parameters ($T_{\rm eff}=4600$K, $\log g=0.8$, $v_{\rm
tur}=2.3~$kms$^{-1}$) for this star. This result suggests that the differences
between the abundances derived by \citet{hill00} and those by this work and
by \citet{preston01}) are mainly due to the atmospheric parameters adopted.

\section{DISCUSSION}\label{sec:disc}

\subsection{Abundance Ratios of Carbon and the Neutron-Capture Elements}\label{sec:abundance}

Figure \ref{fig:cpfe} shows [C/Fe] as a function of [Fe/H] for the
carbon-rich, metal-poor stars studied in the present work, along with those for
other similarly carbon-rich stars described in the literature
\citep{mcwilliam95, norris97a, norris97b, boni99, hill00, aoki01,
norris01}. For comparison, carbon abundances of disk stars and other metal-poor
stars are also shown \citep{ryan91, tomkin92, tomkin95, mcwilliam95,
gustafsson99}. While the metal-rich (disk) stars are dwarfs, the metal-poor
stars include giants that may have been affected by the CN(O)-cycle and
internal mixing in the objects themselves (e.g., by first dredge-up).
Therefore, it remains possible that the original carbon abundances for the most
metal-deficient stars may have been {\it higher} than those shown in the
Figure.

Figure \ref{fig:cpfe} clearly illustrates the large overabundances of carbon,
by more than 1 dex, for three of the stars in the present program.  The excess
of carbon in a fourth star, CS~30314-067, is not distinctive ([C/Fe]$=+0.5$),
but the nitrogen overabundance is large in this star, hence the [C+N/Fe] of
this star is comparable with those of CS~29502-092 and CS~22877-001 (see Table
\ref{tab:res}). Since CS~30314-067 has a lower effective temperature and 
surface gravity than the other stars, the mixing of material affected by
the CN-cycle would tend to reduce [C/Fe] and increase [N/Fe] in the stellar
surface.  Hence we treat CS~30314-067 as a ``carbon-rich'' star similar to
CS~29502-092 and CS~22887-001.  Our analysis confirms the extremely large
overabundance of carbon in CS~22948-027, as found by \citet{hill00}. There is a
large variation in [C/Fe] for stars around [Fe/H]$\sim -3$, and there seems to
be a gap between the very carbon-rich stars ([C/Fe]$\sim$2) and the mildly
carbon-enhanced stars ([C/Fe]$\sim$1), including the newly studied objects
here, as pointed out by \citet{norris97a}.

In Figure \ref{fig:bapfe} we show derived [Ba/Fe], as a function of [Fe/H], for
our carbon-rich stars, together with other metal-poor stars from the
literature.  Carbon-rich stars ([C/Fe]$\geq$0.5) are shown by filled symbols.
The result for CS~22948-027 confirms the large overabundance of barium
found by \citet{hill00}. Four additional stars shown (LP~625-44, LP~706-7, CS~22898-027,
and CS~29497-034) exhibit similar overabundances of barium ([Ba/Fe] $\gtrsim
2$).  

The other four stars in our sample (besides CS~22948-027) exhibit {\it no}
excess of barium. It is known that the barium abundance ratio ([Ba/Fe]) in
metal-poor stars is usually lower than the solar value (i.e., [Ba/Fe] $<0$).
This is often interpreted as a result of the increasing contribution of the
{\it s}-process to the composition of more metal-rich stars
\citep{spite78, truran81}. The abundance ratios of Ba to Eu for metal-poor
stars studied by \citet{mcwilliam98} suggest that the Ba originates in the {\it
r}-process, instead of the {\it s}-process, for most objects with
[Fe/H] $<-2.4$.  From Figure \ref{fig:bapfe}, it is clear that the Ba
abundances of our four stars lie on the general trend of Ba abundances in this
metallicity range. This fact implies that there was likely to have been {\it
almost no} contribution of the {\it s}-process to the Ba abundances in these
carbon-rich stars. In addition to the Ba result, these stars also do not
exhibit overabundances of Sr and Y.  We note that the lower limit of [Ba/Eu] in
CS~30314-067 is higher than the Solar System {\it r}-process value (\S
\ref{ana:metal}), and some additional contribution to heavy elements seems to
exist in this object. Although this could be due to a non-universal
$r$-process, this star also has a low $^{12}$C/$^{13}$C ratio, and the excess
of the Ba/Eu ratio may indicate a contribution from the $s$-process.

\subsection{Carbon-Rich Stars Without Excess of Neutron-Capture Elements}
\label{sec:disc_spoor}

The result of this study is summarized in Figure \ref{fig:bapcn}, in which
[Ba/Fe] is shown as a function of the total abundance ratios of carbon and
nitrogen ([C+N/Fe]). In addition to the abundances of the objects shown in
Figures \ref{fig:cpfe} and \ref{fig:bapfe}, some moderately nitrogen-rich,
Ba-rich objects are also shown in the Figure (HD~25329 and HD~74000 from
Beveridge \& Sneden 1994, and LP~685-47 from Ryan et al. 1991). This Figure
also includes the nitrogen-rich star (CS~22949-037) studied by
\citet{norris01}.  Values of [C+N/Fe] and [Ba/Fe] typical of ``normal''
metal-poor stars are denoted by the cross lower left corner of the Figure.  The
value of [Ba/Fe] $=-0.84$ was adopted from the mean of the Ba abundances of
[Fe/H] $<-2.5$ objects in \citet{mcwilliam98} (excluding the {\it
r}-process-rich star CS~22892-052). The [C+N/Fe] value of ``normal'' metal-poor
stars is not well established, and there appears to be a rather large scatter.
Here we simply assume the solar abundance ratio [C+N/Fe]=0.

The present work indicates that three of our mildly carbon- and/or
nitrogen-rich ([C+N/Fe]$\sim +1$) stars exhibit {\it no} excess of
neutron-capture elements.  This result is difficult to explain by the scenario
of mass transfer from thermally pulsing AGB stars (which probably applies to
CS~22948-027), and requires another process of carbon enrichment that does not
affect the neutron-capture elements.  The carbon and nitrogen overabundances in
the stars with normal neutron-capture elements are moderate, in contrast to the
very large carbon overabundance in the {\it s}-process-rich stars like
CS~22948-027 and LP~625-44 \citep{aoki01}.  However, there exist at least two
extremely carbon- and nitrogen-rich stars with normal neutron-capture element
abundances for their metallicity, CS~22957-027 \citep{norris97b} and
CS~22949-037 \citep{norris01}.  There also exists a moderately carbon-rich and
{\it s}-process-rich star CS~29529-012 \citep{boni99}.  The three moderately
nitrogen-rich, Ba-rich objects shown in Figure \ref{fig:cpfe} may also have an
{\it s}-process-rich nature, as suggested by \citet{beveridge94}.

Discussion of the absolute value of elemental overabundances for the objects
under study is made difficult because, for a given star, it depends on the
assumed dilution due to mixing in the stellar envelope. It may also depend on
the mass accreted from the companion (and the mixing in the donor star), if the
overabundance is caused by mass transfer across a binary system. We simply
point out the importance of the study of chemical composition and binarity for
the class of ``mildly carbon-rich'' stars, as well as for ``extremely
carbon-rich'' objects.

\citet{norris97b} discussed the carbon (and nitrogen) enrichment in
zero metallicity, low-mass stars studied by \citet{hollowell90} as a
candidate for the explanation of the properties of the carbon-rich,
neutron-capture-element-poor star CS~22957-027. The theoretical study
of nucleosynthesis and mixing processes in such stars has been recently
extended to other metallicity and mass ranges by \citet{fujimoto00},
who investigated the possibility of carbon and nitrogen
over-production during He core flash or AGB evolution. They showed
that the low-mass ($M\lesssim 1M_{\odot}$) stars with
$-4\;<\;$[Fe/H]$<\;(-3\;\sim\;-2)$ do not become carbon-rich by core
flash at the tip of the red-giant branch, but rather, through He shell
flashes near the base of the AGB. An important prediction of this
study is that, while the enrichment of {\it s}-process elements occurs
in higher mass ($M\gtrsim 1M_{\odot}$) stars by the third dredge-up
(their case ${\rm II}^{\prime}$), such enrichment does not occur in
low-mass stars (their case II).  Our new results may be examples of this
scenario.

Our present expectation is that an {\it s}-process-rich, carbon-enhanced,
metal-poor star is likely to be the {\it companion} of a higher-mass star from
which neutron-capture-element-rich material has been accreted.  In contrast, a
carbon-rich metal-poor star with normal neutron-capture element abundances
would be a low-mass ($M\sim 0.8M_{\odot}$) star in which carbon and nitrogen
have been {\it self-enhanced} during its AGB evolution without affecting the
abundances of its neutron-capture elements.  This latter, ``intrinsic scenario,''
would require that we are presently seeing such stars during their brief
lifetime on the AGB.  An alternative possibility is that these stars are
companions of slightly higher-mass ($0.8M_{\odot} < M\lesssim 1M_{\odot}$)
stars from which carbon-rich material without excess of neutron-capture
elements has been accreted. It should be noted that the binarity of the
carbon-rich, neutron-capture element-poor star CS~22957-027 was recently
confirmed by \citet{preston01}.  It is clearly of great importance to search
for binarity amongst other similar stars.

\citet{fujimoto00} also predicted overabundances of nitrogen due to hydrogen
mixing following the helium shell flash in low-mass, low- metallicity
stars. Two of our stars with normal neutron-capture-element abundances exhibit
large nitrogen abundances (CS~30314-067 and CS~29502-092, [N/H]$\sim +1$), but
one does not (CS~22877-001, [N/H]$\sim 0$). As mentioned in \S
\ref{sec:cn}, the systematic uncertainty of the nitrogen abundance derived from
the CN molecular band is large, but the relative abundances among our stars
should be reliable. A variation of nitrogen abundances certainly exists among
our mildly carbon-rich stars with similar metallicity ([Fe/H]$\sim -2.8$). This
result provides a constraint on the evolutionary models of very metal-poor AGB
stars.

The above discussion concerning carbon and nitrogen production by low-mass
metal-poor stars is still quite primitive, and further tests of consistency
between observation and models are indispensable. We also need to consider
whether other possible processes that enrich carbon at very low metallicity
exist.  There is no attractive candidate at present, but we should recall the
discussion by \citet{norris97a} concerning the possibility of large amounts of
carbon and nitrogen ejection from ``hypernovae'' of very high mass in the early
era of the Galaxy. As yet there has been no quantitative analysis of the
role of mass loss in driving freshly-synthesized C and/or N into the
interstellar medium from the surface of high-mass, low-metallicity stars. If
such stars eventually collapse to black holes as failed supernovae, but have
already expelled significant quantities of C or N, they would enrich the early
protogalaxy in these elements without also yielding heavier metals. Such
objects could contribute to the high [C/Fe] and [N/Fe] ratios observed in some
of the oldest objects. A challenge for this mechanism is that stellar winds are
reduced in low metallicity environments, but massive stars that not only attain
high luminosities but also synthesize their own C and/or N may overcome this
potential problem.  We hope this mechanism will be explored quantitatively in
coming years.

\subsection{Carbon Enrichment in the {\it r}-Process-Enhanced star
CS~22892-052}\label{sec:disc_r}

The discovery of the extremely {\it r}-process-enhanced metal-poor star
CS~22892-052 has had a major impact on the study of neutron-capture
nucleosythesis and cosmochronology using the radioactive element Th. The excess
of carbon in this object ([C/Fe] $\sim +1.0$) has influenced the subsequent
discussion concerning the site of the {\it r}-process which produced the
abundance pattern in this star, hence is worthy of consideration in some
detail.

Figure \ref{fig:bapcn} shows that CS~22892-052 is also mildly
carbon-rich (and nitrogen-rich), similar to three of our stars.  Our
mildly carbon-rich stars have a quite ``normal'' abundance pattern
except for carbon and nitrogen, in contrast to {\it s}-process-rich
stars like LP~625-44 \citep{aoki00}. If another process provides
CS~22892-052 with {\it r}-process-rich material, an almost pure {\it
r}-process abundance pattern will be observed in the object. Mildly carbon-rich stars seem common amongst very metal-poor
giants. It is possible that an independent, infrequent
process must be invoked to produce the {\it r}-process element
overabundances even in stars with carbon and nitrogen excesses like
CS~22892-052.  One possibility is the ``peppering'' of the surface of
CS~22892-052 by the nucleosynthesis products from a massive companion
that underwent a Type II supernova explosion, and may now be present
as a massive collapsed object, such as a black hole (Qian \&
Wasserburg 2001).  The suggestion by Preston \& Sneden (2001) that
CS~22892-052 may exhibit an anomalously short period ($\sim 180$ days)
might also be expected in such a scenario.  Clearly, this period must
be confirmed by additional data, and searches for periodicity in the
other known extremely {\it r}-process-enhanced metal-poor star,
CS~31082-001 (Cayrel et al. 2001) should be undertaken. In any event,
the lack of neutron-capture element enhancement in our mildly
carbon-rich stars certainly suggests that the origin of {\it
r}-process-element enhancement in CS~22892-052 might be {\it
independent} of the process responsible for its carbon and nitrogen
excesses.  It should be noted that neither the moderately {\it
r}-process-rich star HD~115444 \citep{westin00}, nor the extremely
{\it r}-process-rich star CS~31082-001, exhibit carbon enhancement
\citep{cayrel01}.

\subsection{Isotope Ratios of Carbon and Nitrogen}

The species $^{13}$C and $^{14}$N are both produced from $^{12}$C via
hydrogen burning in the CN cycle, hence the measurement of $^{13}$C and N in
several of our stars permits us to compare the isotope ratios to those
expected.  Although we do not have isotope information for N, it is reasonable
to assume that the dominant isotope is $^{14}$N.

When the CN cycle runs in equilibrium, the isotope ratios attain values
$^{12}$C/$^{13}$C $\simeq$ 3 and $^{12}$C/$^{14}$N $\simeq$ 0.03, equivalent to
$^{13}$C/$^{14}$N $\simeq$ 0.01 \citep{arnould99}.  When we measure $^{12}$C,
we do not know what fraction has passed through the CN cycle, so the isotope
ratios involving $^{12}$C are difficult to interpret. However, since both
$^{13}$C and $^{14}$N are believed to be produced only by hydrogen burning in
the CN-cycle sequence, their isotope ratios should reflect the degree of
processing.

In Table \ref{tab:iso} we show the element and isotope ratios for the four
stars in this paper with either measured $^{13}$C values or limits, and the two
stars studied by Norris et al. (1997a,b) and Aoki et al. (2001). It is
immediately clear that the $^{13}$C/$^{14}$N ratio is well in excess of the
CN-cycle equilibrium value. That is, even when $^{12}$C is hydrogen burning to
$^{13}$C, it is not being burnt on to $^{14}$N in the equilibrium ratio. It is
perhaps surprising that this is the case even in the star with
$^{12}$C/$^{13}$C = 5.

The non-equilibrium isotope ratios should help identify the sites in which such
large carbon and nitrogen excesses were produced. Normally, the CN cycle reaches
equilibrium on a short timescale, after a few proton-capture lifetimes,
equivalent to the consumption of typically one proton per initial nucleus
(e.g., Clayton 1968). However, one case in which high $^{13}$C/$^{14}$N ratios
might be achieved is if the star's evolutionary timescale during CN cycling is
shorter than the timescale for the $^{13}$C(p,$\gamma$)$^{14}$N reaction, so
conditions become unfavorable for hydrogen burning before $^{14}$N production
completes.  This might imply that the hydrogen burning occurred during more
rapid stages of evolution, on the giant branch or during short-lived thermal
pulses on the AGB, rather than on the main-sequence.  Truncation of the CN
cycle could have happened if the temperature fell below that required for
hydrogen burning, or if protons were in short supply, as might happen in the He
intershell of a thermally-pulsing AGB star. The latter possibility is
underlined by the analysis of the very-metal-poor binary LP~625-44: the
$^{13}$C pocket of its AGB component, produced by protons mixed down from the
H-rich envelope, was only 1/24th that expected from higher-metallicity models
(Ryan et al. 2001).  Proton starvation could leave a star unable to burn beyond
$^{13}$C, and lead to a truncated CN cycle.

It was noted above that a high, non-equilibrium $^{13}$C/$^{14}$N ratio exists
even for the star with the low, equilibrium value of $^{12}$C/$^{13}$C = 5.
However, this may nevertheless be compatible with a truncated CN cycle.
Clayton (1968, his Fig. 5-15) shows the approach to equilibrium for burning at
20$\times 10^6$~K. At a stage when 0.8--0.9 protons have been consumed per
initial nucleus, before the cycle reaches equilibrium, $^{13}$C/$^{14}$N $\sim
$ 0.5, but already $^{12}$C/$^{13}$C $\simeq$ 4; $^{13}$C and $^{12}$C come
into equilibrium with each other faster than with $^{14}$N. $^{13}$C/$^{14}$N
then decreases as equilibrium is approached, but with little change in
$^{12}$C/$^{13}$C.  Hence, the apparent contradiction between equilibrium
$^{12}$C/$^{13}$C ratios and non-equilibrium $^{13}$C/$^{14}$N ratios may be a
pointer to the time at which proton truncation occurred, as well as to the mean
number of proton captures.  The timescale for the CN cycle to reach this stage
in Clayton's example was $\sim 10^4$~yr, of the same order as the interval
between the thermal pulses of an AGB star.  Furthermore, the equilibrium
$^{13}$C/$^{14}$N ratio is temperature dependent, and would be higher for $T >
20\times 10^6$~K, making it even easier to avoid overproduction of $^{14}$N
relative to $^{13}$C in hotter environments such as an AGB star's He
intershell.  We would welcome a more thorough testing of these possibilities in
AGB and other stellar models, to see whether models can be identified that
possess the conditions required to reproduce the observed ratios.

\section{SUMMARY}

Our abundance analysis for very metal-poor stars with strong CH bands, based on
high-resolution spectra, revealed that four of our five program stars do not
exhibit overabundances of barium ([Ba/Fe]$<0$).  Three of the four objects
without Ba excesses have mildly enhanced carbon and/or nitrogen
([C+N]$\sim$+1).  Prior to this work, only one carbon-rich,
neutron-capture-element-poor star was known (CS~22957-027). Our study indicates
that this kind of object is not uncommon among very metal-deficient stars.

We have discussed possible processes to explain the chemical nature of
carbon-rich, metal-poor stars. One candidate is the helium shell flash near the
base of the AGB in low-mass ($M\lesssim 1M_{\odot}$) stars proposed by
\citet{fujimoto00}. They showed that carbon and nitrogen enrichment occurs by
this process in stars with $-4 \lesssim$ [Fe/H] $\lesssim -3(\sim-2)$ stars,
but the {\it s}-process elements are enhanced by the third dredge-up only in
higher mass ($M \gtrsim 1M_{\odot}$) stars. Our result may be interpreted
within this framework, but a further test of consistency between observations
and the models, such as for nitrogen abundances, is required.

The fact that mildly carbon-rich stars with normal neutron-capture-element
abundances are not rare amongst metal-poor giants suggests that the moderate
overabundance of carbon ([C/Fe]$\sim$1) in the famous {\it r}-process-enhanced
star CS~22892-052 may have originated in the same process that produced our
carbon-rich stars.  This implies a decoupling of carbon and {\it r}-process
production processes for this object.

Carbon-enhanced stars are quite numerous among very metal-deficient stars.
Nevertheless, among the limited number of such stars studied so far, a large
variation in the abundances of carbon and neutron-capture elements has been
found. To understand the nature of this important class of objects, and their
role in the early Galaxy, detailed abundance studies for carbon-rich objects
covering a wider metallicity range is desirable.  Long-term studies to detect,
and quantify, the existence of binarity among such stars are crucial as well.

\acknowledgments 

The authors are grateful to the Australian Time Allocation Committee for their
continued support for our studies of the most metal-deficient stars and to the
Director and staff of the Anglo-Australian Observatory for providing the
facilities for this study. The authors wish to thank Dr. John Lattanzio and Dr.
M. Fujimoto for helpful discussions. Thanks are also due to the anonymous
referee for useful comments. T.C.B. acknowledges partial support from NSF grant
AST 00-98549. S.G.R acknowledges support from PPARC grant PPA/O/S/1998/00658.

\clearpage
\begin{deluxetable}{p{23mm}p{4mm}cp{5mm}p{20mm}rp{5mm}p{7mm}p{24mm}l}
\tablewidth{0pt}
\tablecaption{PROGRAM STARS AND OBSERVATIONS \label{tab:obs}}
\startdata
\tableline
\tableline
Star & $V$ & $B-V$ & $GP$ & Wavelength & Exp.$^{\ast}$ & S/N & $\sigma_{\small\rm W}$ & ~ ~ $V_{\rm rad}$ & Obs. Date (JD) \\ 
 & & &({\AA}) & ~ ~  ({\AA}) & (min) & & (m{\AA}) & ~ ~ (kms$^{\small -1}$) & \\ 
(1)&(2)&(3)&(4)& ~ ~ (5)&(6) ~&(7)&(8)& ~ ~ (9)& ~ ~ ~ (10) \\
\tableline
CS~22948-027 & 12.66 & 1.13 & 7.79 & 5010-8530 &  75 (3) & 51 & 6.7 & $-63.33\pm0.56$ & June 1994 (2449520) \\
CS~30314-067 & 11.85 & 1.13 & 5.80 & 3740-4800 & 120 (4) & 77 & 3.9 &                 & Aug. 1997 \\
            &       &      &       & 4970-8400 &  60 (2) & 95 & 4.0 & $+145.58\pm0.29$ & Aug. 1997 (2450684) \\
CS~22877-001 & 12.20 & 0.73 & 4.54 & 3620-4880 & 120 (4) & 36 & 5.6 &                 & Apr. 1998 \\
            &       &      &       & 4400-7710 &  60 (2) & 77 & 4.6 & $+166.12\pm0.25$ & Apr. 1998 (2450914) \\
CS~30312-100 & 12.95 & 0.59 & 3.49 & 4440-7710 & 120 (4) & 57 & 5.1 & $-128.91\pm0.28$ & Apr. 1998 (2450914) \\
CS~29502-092 & 11.88 & 0.77 & 4.51 & 3740-4800 & 150 (5) & 70 & 4.1 &                 & Aug. 1998 \\
            &       &      &       & 4600-7130 &  60 (2) & 73 & 4.0 & $-67.35\pm0.27$ & Aug. 1998 (2451039) \\
\tableline
\enddata
~ \\

$\ast$ Number of exposures is given in the parentheses. 

\end{deluxetable}


\begin{deluxetable}{p{12mm}p{8mm}p{8mm}cccccl}
\tablecaption{EQUIVALENT WIDTHS FOR PROGRAM STARS (m{\AA})\label{tab:ew}}
\tablewidth{0cm}
\startdata
\tableline
\tableline
$\lambda$({\AA}) & $\chi$(eV) & $\log gf$ & 30314-067 & 29502-092 & 22877-001 & 30312-100 & 22948-027 & $gf$ Ref.  \\
\tableline
  Na I:      &&&&&&&&\\
   5889.95 &  0.000 &  0.11 &  165 &  135 &  118 &  158 &  209 & (1) \\
   5895.92 &  0.000 &  $-$0.19 &  153 &   97 &   95 &  130 &  195  & (1)\\
  Mg I:     &&&&&&&&\\
   3829.35 &  2.710 &  $-$0.48 &  197 &   ...&  133 &   ...&   ... & (2)$-$0.25\\
   3832.30 &  2.710 &  $-$0.13 &  215 &  169 &  173 &   ...&   ... & (2)$-$0.25\\
   3838.30 &  2.720 &  $-$0.10 &  203 &  181 &  218 &   ...&   ... & (3)$-$1.05\\
\enddata
\end{deluxetable}

\begin{deluxetable}{llllccl}
\tablewidth{0pt}
\tablecaption{STELLAR PARAMETERS \label{tab:param}}
\startdata
\tableline
\tableline
Star & $T_{\rm eff}$(K) & $\log g$ & $v_{\rm tur}$(kms$^{\small -1}$) & [Fe/H] & [C/H] & [N/H] \\ 
\tableline
CS~22948-027 & 4600 & 1.0 & 2.2 & $-$2.57 & $-$0.6 & $-$0.8 \\
CS~30314-067 & 4400 & 0.7 & 2.5 & $-$2.85 & $-$2.4 & $-$1.7 \\
CS~22877-001 & 5100 & 2.2 & 2.0 & $-$2.72 & $-$1.7 & $-$2.7 \\
CS~30312-100 & 5500 & 3.0 & 2.3 & $-$2.17 & - & - \\
CS~29502-092 & 5000 & 2.1 & 1.8 & $-$2.76 & $-$1.8 & $-$2.1 \\
\tableline
\enddata
\end{deluxetable}

\begin{deluxetable}{@{}l@{\extracolsep{\fill}}c@{\extracolsep{\fill}}c@{\extracolsep{\fill}}c@{\extracolsep{\fill}}c@{\extracolsep{\fill}}c@{\extracolsep{\fill}}c@{\extracolsep{\fill}}c@{\extracolsep{\fill}}c@{\extracolsep{\fill}}c@{\extracolsep{\fill}}c@{\extracolsep{\fill}}c@{\extracolsep{\fill}}c@{\extracolsep{\fill}}c@{\extracolsep{\fill}}c@{\extracolsep{\fill}}c@{\extracolsep{\fill}}c@{\extracolsep{\fill}}c@{\extracolsep{\fill}}c@{\extracolsep{\fill}}c@{\extracolsep{\fill}}c@{\extracolsep{\fill}}c}
\tablecaption{[FE/H] AND RELATIVE ABUNDANCE, [X/FE] \label{tab:res}}
\startdata
\tableline
\tableline
     &\multicolumn{3}{c}{CS~30314-067} &&\multicolumn{3}{c}{CS~29502-092}   && \multicolumn{3}{c}{CS~22877-001}  && \multicolumn{3}{c}{CS~30312-100}  && \multicolumn{3}{c}{CS~22948-027} \\ 
  \cline{2-4}  \cline{6-8}    \cline{10-12}   \cline{14-16}    \cline{18-20} \\
     & [X/Fe] & n & s.e.  && [X/Fe] & n & s.e.  && [X/Fe] & n & s.e.  && [X/Fe] & n & s.e.  && [X/Fe] & n & s.e. \\
\noalign{\smallskip}
\hline
\noalign{\smallskip}
C     & $+$0.5  &   & 0.18 && $+$1.0  &   & 0.18 && $+$1.0  &   & 0.18 && -       &   &      && $+$2.0  &   & 0.18  \\ 
N     & $+$1.2  &   & 0.24 && $+$0.7  &   & 0.24 && 0.0     &   & 0.24 && -       &   &      && $+$1.8  &   & 0.24  \\ 
C+N   & $+$0.8  &   &      && $+$1.0  &   &      && $+$0.9  &   &      && -       &   &      && $+$2.0  &   &      \\ 
\ion{Na}{1}  & $-$0.08 & 2 & 0.18 && $-$0.01 & 2 & 0.18 && $-$0.24 & 2 & 0.13 && $+$0.04 & 2 & 0.12 && $+$0.68 & 2 & 0.20  \\ 
\ion{Mg}{1}  & $+$0.42 & 5 & 0.10 && $+$0.37 & 7 & 0.07 && $+$0.29 & 7 & 0.08 && $+$0.17 & 3 & 0.10 && $+$0.59 & 2 & 0.11  \\
\enddata

~ \\

$\ast$ The upper-limit of Li is given for $\log\epsilon_{\rm Li}$ instead of [Li/Fe].

\end{deluxetable}

\begin{deluxetable}{@{}l@{\extracolsep{\fill}}c@{\extracolsep{\fill}}c@{\extracolsep{\fill}}c@{\extracolsep{\fill}}c@{\extracolsep{\fill}}c@{\extracolsep{\fill}}c@{\extracolsep{\fill}}c@{\extracolsep{\fill}}c@{\extracolsep{\fill}}c@{\extracolsep{\fill}}c}
\tablecaption{ERROR ESTIMATES FOR [FE/H] AND [X/FE] \label{tab:error}}
\startdata
\tableline
\tableline
     &\multicolumn{4}{c}{CS~30314-067 (giant)} &&\multicolumn{4}{c}{CS~30312-100 (subgiant)} \\ 
  \cline{3-6}  \cline{8-11} \\
   & & $\Delta T_{\rm eff}$ &  $\Delta \log g$ & $\Delta v_{\rm tur}$  & $\sigma_{\rm random}$ & & $\Delta T_{\rm eff}$ &  $\Delta \log g$ & $\Delta v_{\rm tur}$  & $\sigma_{\rm random}$ \\ 
   & & $+100$K &  $+0.3$ & $+0.5$kms$^{-1}$  &  & & $+100$K &  $+0.3$ & $+0.5$kms$^{-1}$  & \\
\noalign{\smallskip}
\hline
\noalign{\smallskip}
C           & &   +0.00 & $-$0.06 &   +0.09 & 0.14  & &         &         &         &        \\
N           & &   +0.00 & $-$0.03 &   +0.09 & 0.22  & &         &         &         &        \\
\ion{Na}{1} & &   +0.01 & $-$0.03 & $-$0.16 & 0.075 & &         &         & $-$0.08 & 0.075  \\
\ion{Mg}{1} & &   +0.01 & $-$0.04 & $-$0.08 & 0.051 & & $-$0.03 & $-$0.02 & $-$0.07 & 0.067  \\
\enddata
\end{deluxetable}

\begin{deluxetable}{lccccccc}
\tablewidth{0pt}
\tablecaption{ISOTOPE RATIOS OF CARBON AND NITROGEN \label{tab:iso}}
\startdata
\tableline
\tableline
Star        &[Fe/H]&[C/Fe]&[N/Fe]&$^{12}$C/$^{13}$C&$^{12}$C/N&$^{13}$C/N&Ref.\\
\tableline
CS~30314-067&--2.85& 0.5  & 1.2  &   5             & 0.63     & 0.13     &1   \\
CS~29502-092&--2.76& 1.0  & 0.7  &  20             & 7.2      & 0.36     &1   \\
CS~22948-027&--2.57& 2.0  & 1.8  &  10             & 5.5      & 0.55     &1   \\
CS~22877-001&--2.17& 1.0  & 0.0  &$>10$            & 36       &$<3.5$    &1   \\
CS~22957-027&--3.38& 2.2  & 2.0  &  10             & 5.5      & 0.55     &2   \\
LP~625-44   &--2.68& 1.95 & 1.65 &  20             & 7.2      & 0.36     &3,4 \\
\tableline
\enddata

~\\

References: (1)This work; (2)Norris et al. (1997b); (3)Norris et al. (1997a); (4)Aoki et al. (2001)
\end{deluxetable}

\clearpage
\begin{figure}	
\caption[]{Observed spectra near the \ion{Ba}{2} $\lambda 4554$
line. The spectra of two giants (CS~30314-067 and CS~29502-092) are
compared with that of the {\it s}-process-rich subgiant LP~625-44
\citep{aoki00}. The \ion{Ba}{2} $\lambda 4554$ lines in the two giants
are much weaker than that in LP~625-44. No lines of other neutron-capture
elements are detected in the two giants, while numerous lines
of \ion{La}{2}, \ion{Ce}{2}, \ion{Nd}{2} and \ion{Sm}{2} appear in
LP~625-44.
}
\label{fig:sp4554}
\end{figure}
\begin{figure}
\caption[]{Predicted colors ($B-V$ and $R-I$) calculated using synthetic
spectra (see text). The colors calculated by \citet{hill00} are
plotted for comparison.}
\label{fig:teff}

\caption[]{Comparison of the observed spectra (dots) with the
synthetic ones (lines) in the region near the CN band at 3883{\AA} and
CH band at 4323{\AA}. On the right the three synthetic spectra differ
in steps of $\Delta$[C/Fe] = 0.3, while on the left the computations
were done with the best-fit [C/Fe] determined at 4323{\AA} and steps
of $\Delta$[N/Fe] = 0.3. The adopted values of [C/Fe] and [N/Fe],
which produce the central synthetic spectrum in each panel, are given
in the figures.}

\label{fig:chcn}
\end{figure}
\begin{figure}
\caption[]{Comparison of the observed spectra (dots) with the synthetic
ones (lines) in the region near the C$_{2}$ bands at 5160{\AA} (top and
middle) and at 5635{\AA} (bottom). The computations were done with steps
of $\Delta$[C/Fe] = 0.3.}
\label{fig:c2}
\end{figure}
\begin{figure}
\caption[]{Comparison of the observed spectra (dots) with the synthetic
ones (lines) around 4220{\AA} for CS~30314-067 (upper three panels) and
CS~29502-092 (lower three panels). The carbon isotope ratios assumed in
the computations are $^{12}$C/$^{13}$C=3, 5, and 10 for CS~30314-067
and $^{12}$C/$^{13}$C=10, 20, and 40 for CS~29502-092. 
}
\label{fig:c1213}
\end{figure}
\begin{figure}
\caption[]{Carbon abundance ratios ([C/Fe]) as a function of [Fe/H].
Present results are shown by stars, and compared with results from
other studies (see text).
}
\label{fig:cpfe}
\end{figure}
\begin{figure}
\caption[]{Present results for barium abundances ([Ba/Fe]) are shown by
stars. The barium abundances of metal-poor stars derived by other
studies are also shown.  The line connecting two symbols is for the star
CS~22948-027, also studied by Hill et al. (2000).  
}
\label{fig:bapfe}
\end{figure}
\begin{figure}
\caption[]{Barium abundances as a function of the total abundance
ratios of carbon and nitrogen ([(C+N)/Fe]) for carbon or nitrogen-rich
stars.  The abundances of HD~122563, which is not a carbon-rich star,
are displayed for comparison. 
}
\label{fig:bapcn}
\end{figure}

\plotone{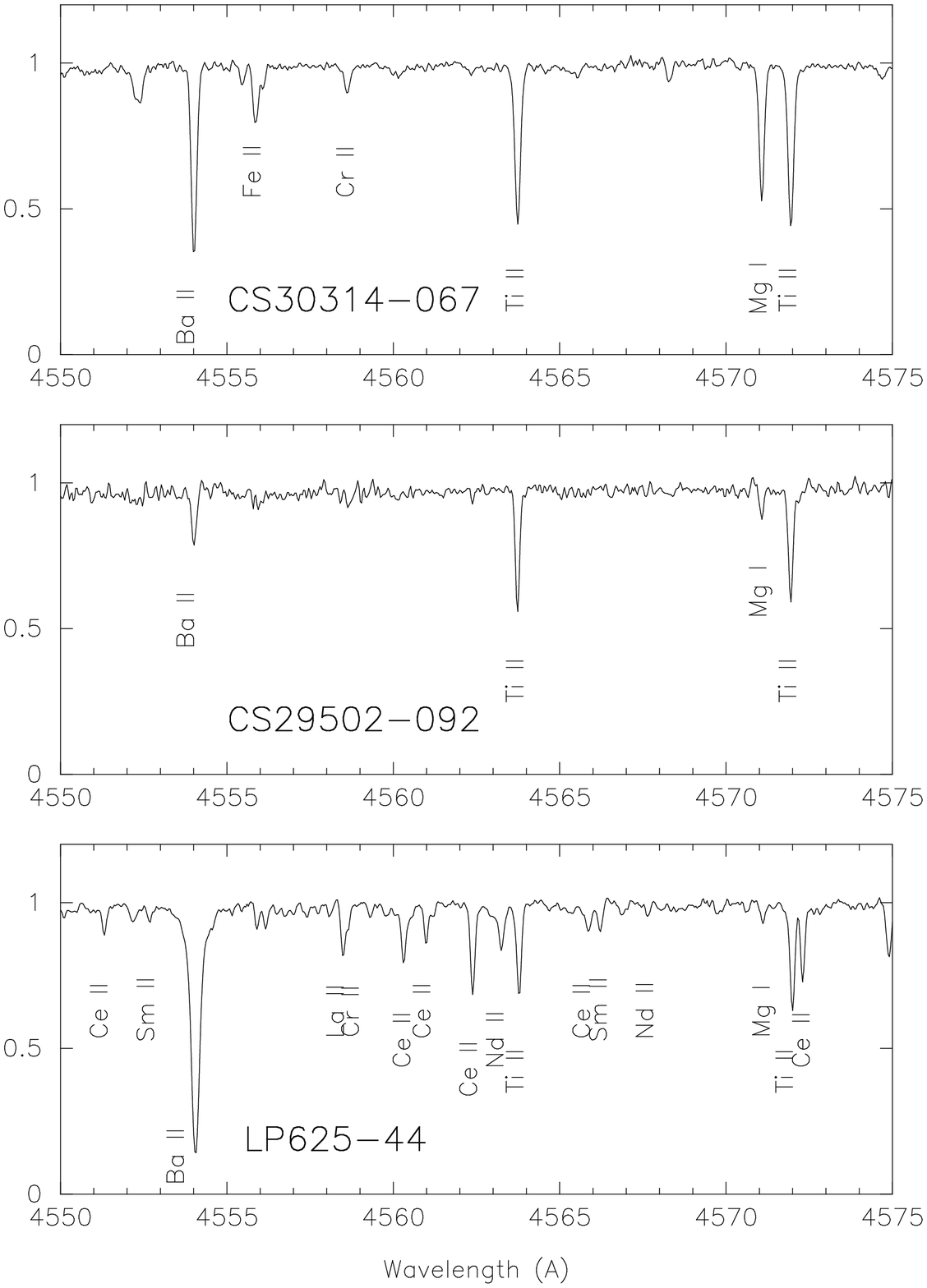}
\newpage
\plotone{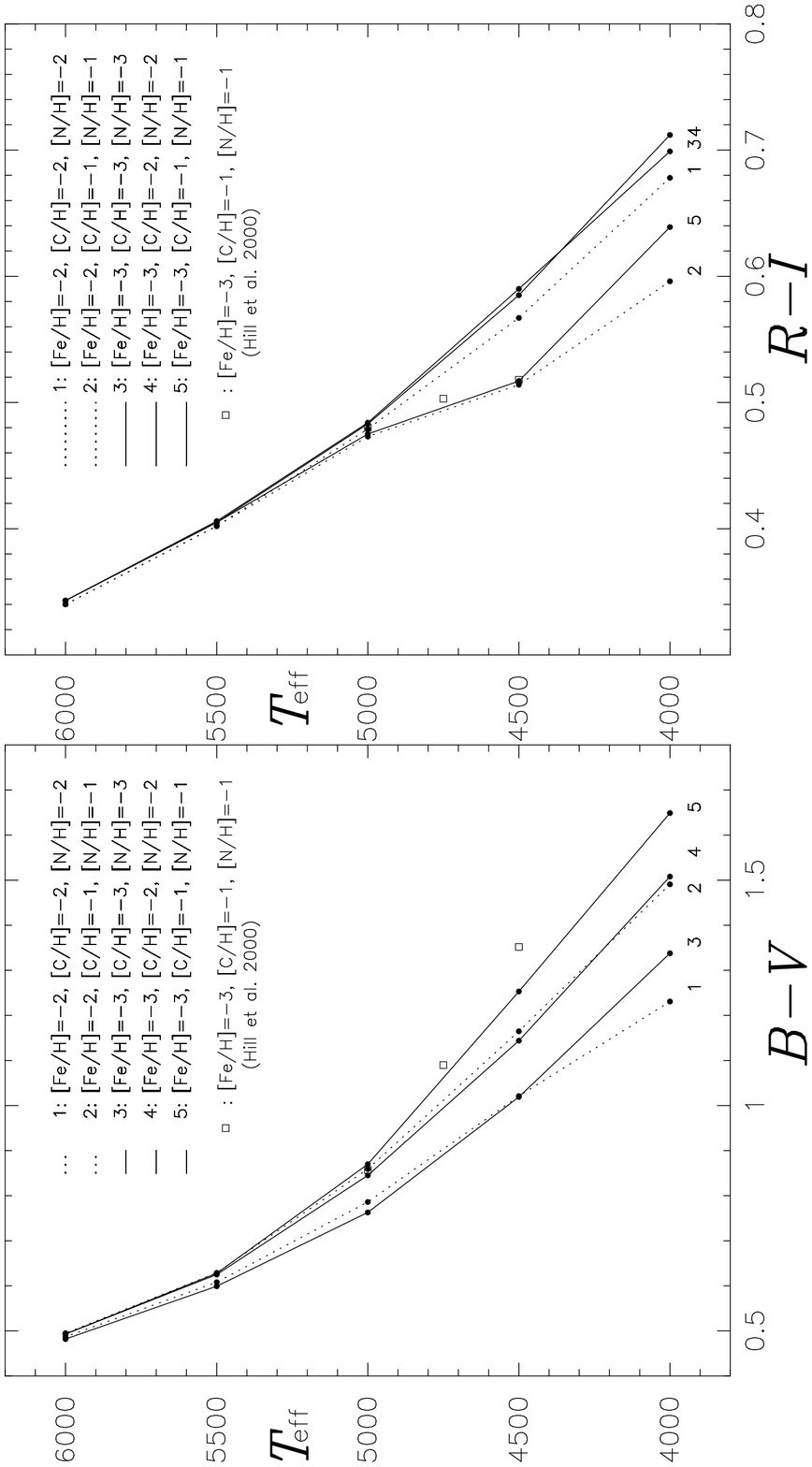}
\newpage
\plotone{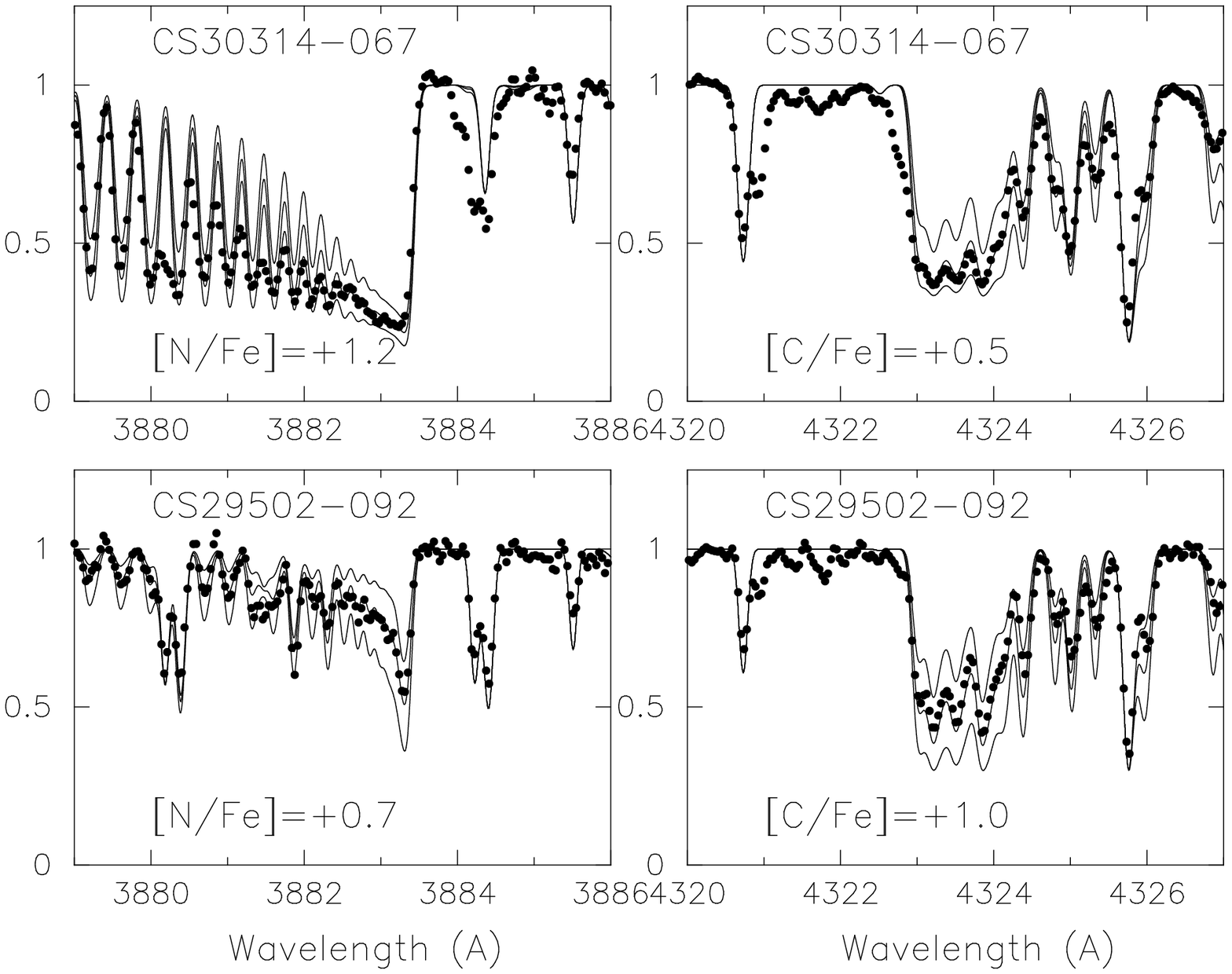}
\newpage
\plotone{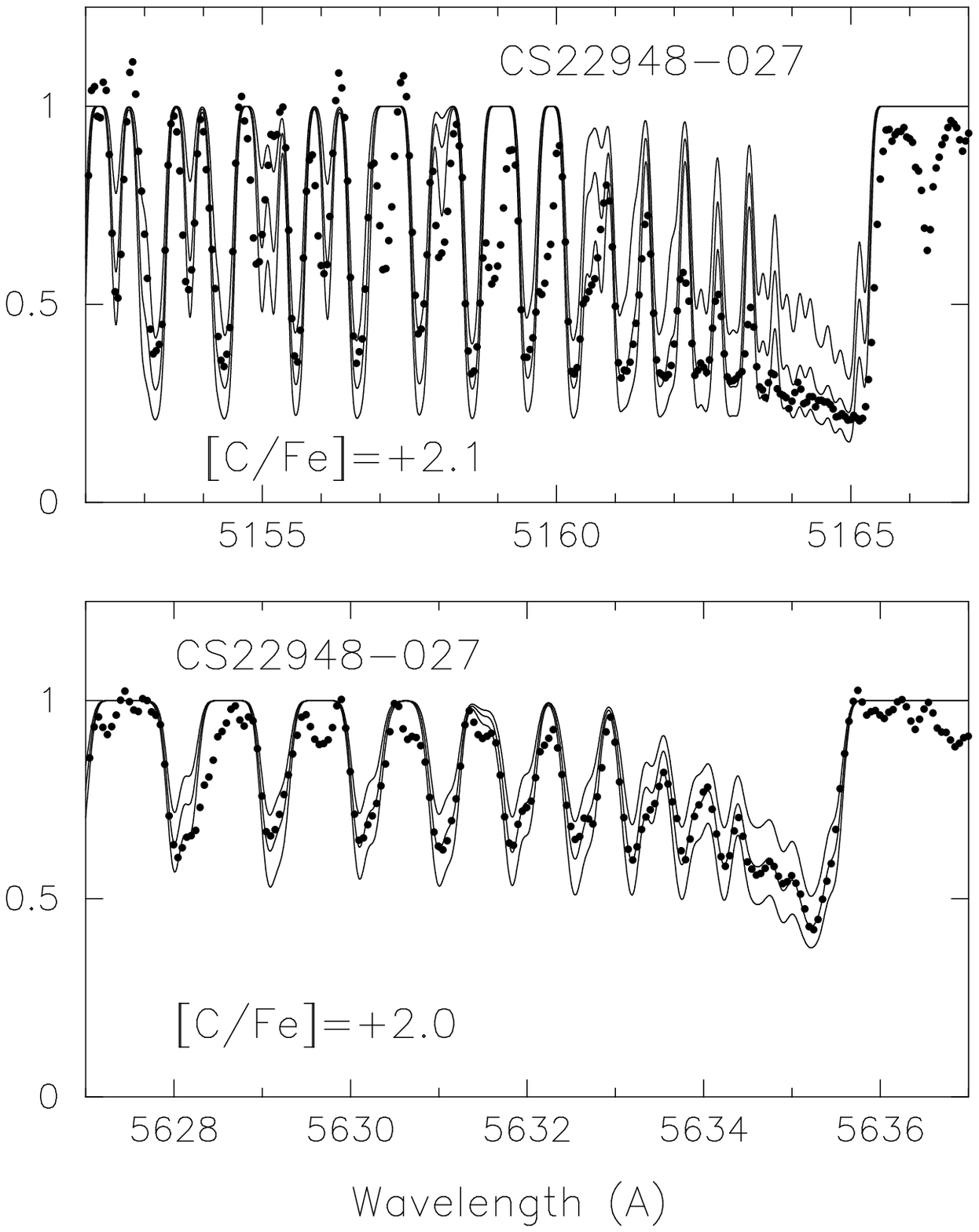}
\newpage
\plotone{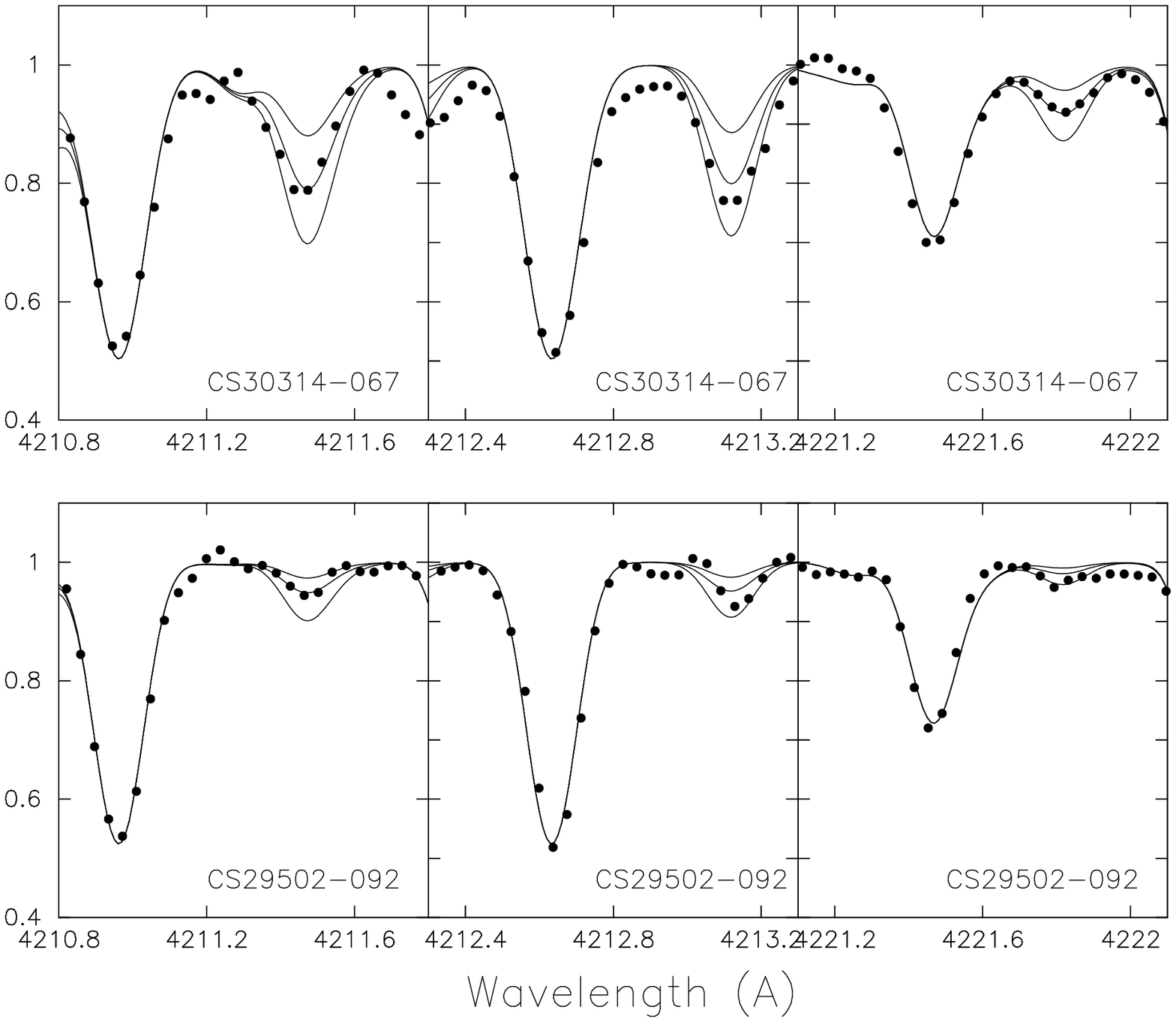}
\newpage
\plotone{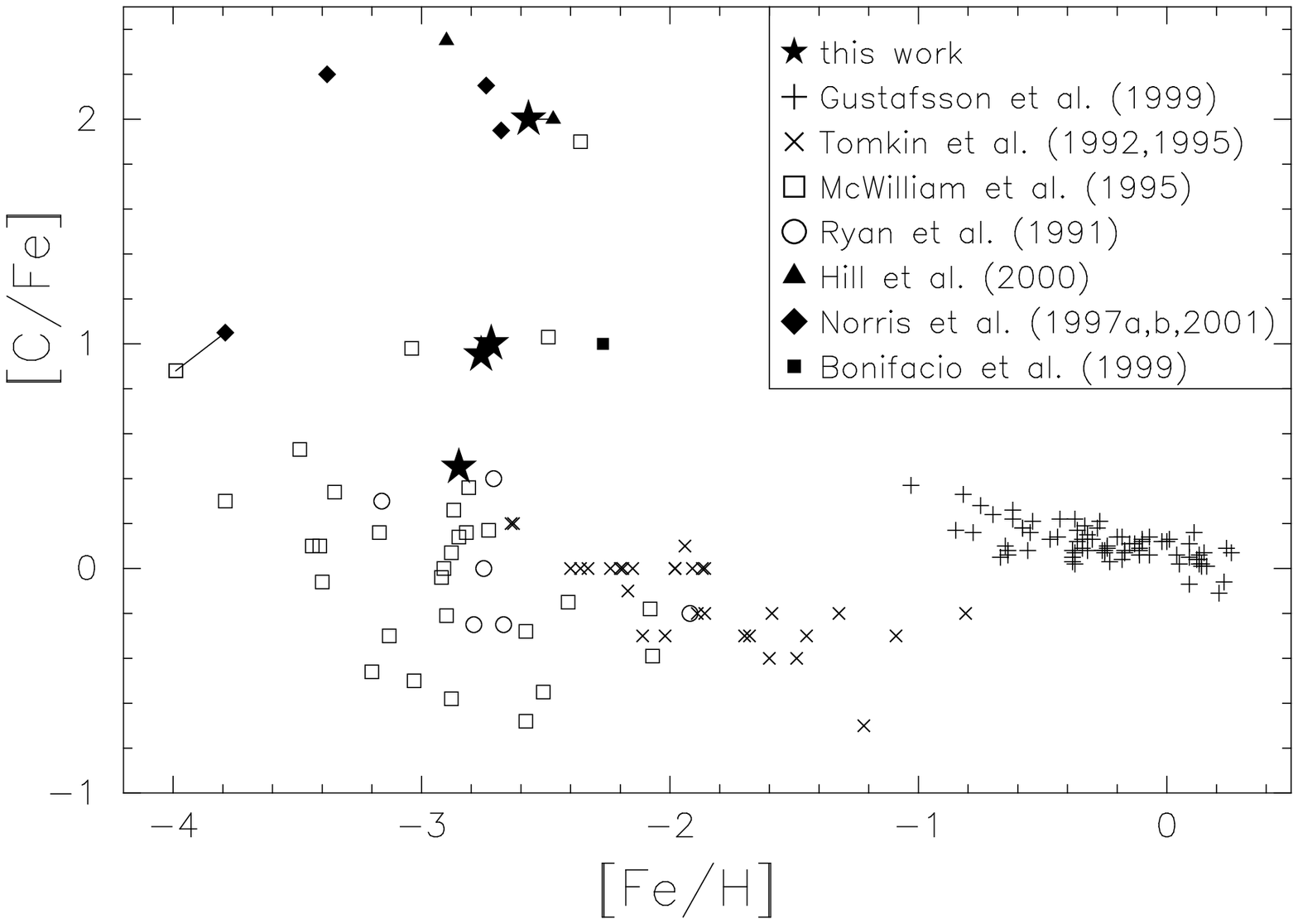}
\newpage
\plotone{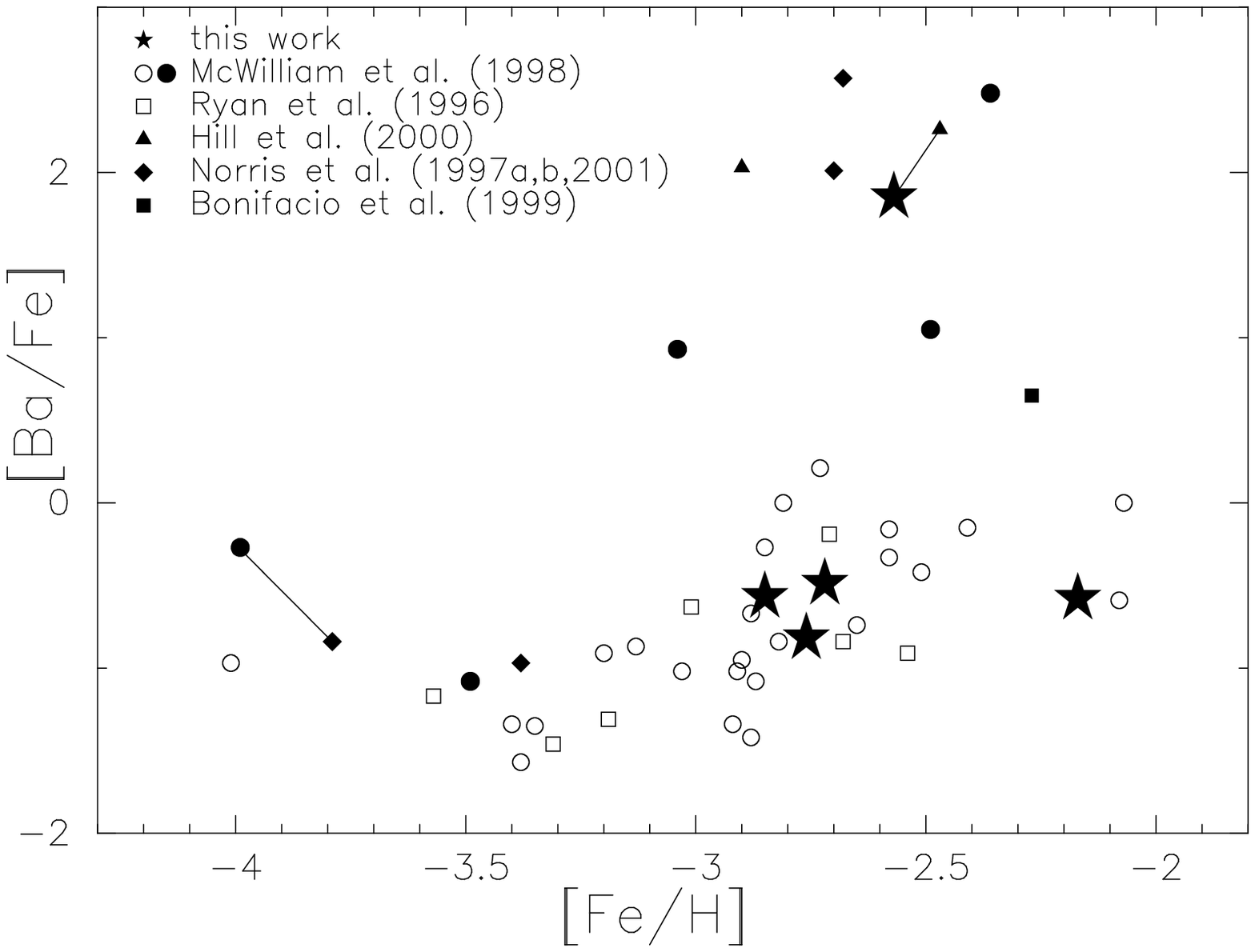}
\newpage
\plotone{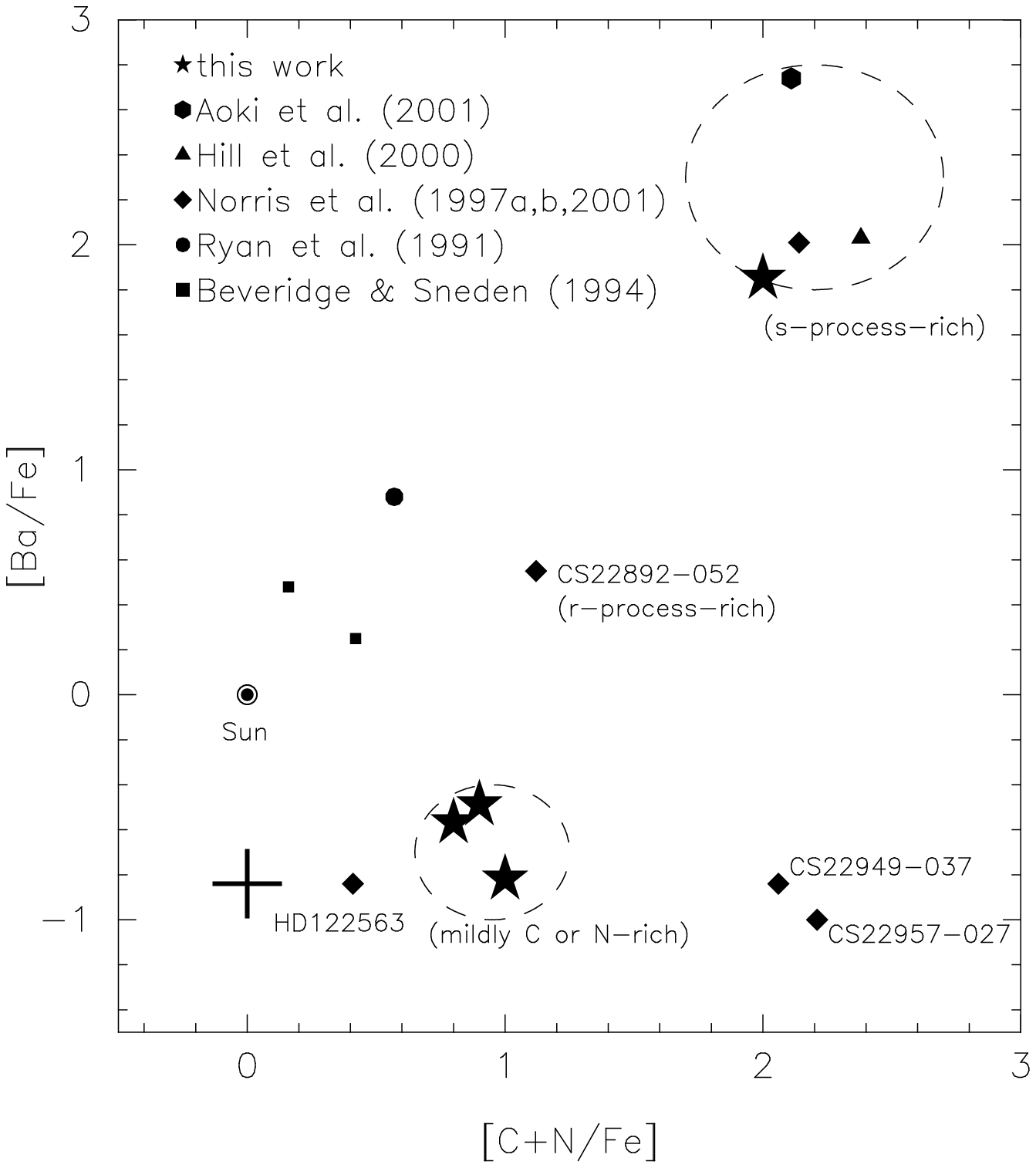}
\newpage

\end{document}